\documentclass{aa} 

\usepackage{txfonts}
\usepackage{natbib}
\usepackage{graphicx}
\begin{document} 
  
  
\title{The Andromeda Project. I. Deep HST-WFPC2 V,I photometry of 16 fields   
    toward the disk and the halo of the M31 galaxy. \\  
    Probing the stellar content and metallicity distribution.
\thanks{Based on observations made with the NASA/ESA Hubble Space Telescope, 
obtained at the Space Telescope Science Institute, which is operated by the
Association of Universities for Research in As\-tronomy, Inc., under
NASA contract NAS 5-2655.  These observations are associated with 
proposal GO-6671.} } 
  
 \author{M. Bellazzini\inst{1} 
       \and C. Cacciari\inst{1}
       \and L. Federici\inst{1}
       \and F. Fusi Pecci\inst{1}
       \and M. Rich\inst{2}}

\offprints{M. Bellazzini, \email{bellazzini@bo.astro.it}} 

\institute{Osservatorio Astronomico di Bologna, Via Ranzani 1, 40127, Bologna, 
ITALY
 \and Dept. of Physics and Astronomy, Division of Astronomy and 
Astrophysics, University of California, Los Angeles, CA 90095-1562} 

  
\date{Received {}; accepted {}} 
  
  
  
\abstract{    
   
We have obtained HST-WFPC2 F555W and F814W photometry for 16 fields in the   
vicinity of the luminous nearby spiral galaxy     
M31, sampling the stellar content of the disk and the halo      
at different distances from the center, from $\sim$ 20 to $\sim$ 150     
arcmin (i.e. $\sim$ 4.5 to 35 kpc), down to limiting V and I magnitudes     
of $\sim$ 27.     
   
 The Color-Magnitude diagrams (CMD) obtained for each field show the presence of    
complex stellar populations, including an intermediate age/young population     
and older populations with a wide range of metallicity.     
Those fields superposed on the disk of M31 generally show   
a blue plume of stars which we identify with main sequence members.   
According to this interpretation, we find that the star formation rate   
over the   
last 0.5 Gyr has varied dramatically with location in the disk.     
   
The most evident feature of all the CMDs is a prominent Red Giant Branch    
(RGB) with a descending tip in the V band,    
characteristic of metallicity higher than 1/10 Solar.   
A red clump is clearly detected in all of the fields, and a weak blue   
horizontal branch is frequently present.   
   
The metallicity distributions, obtained by comparison of the RGB stars    
with globular cluster templates, all show a long, albeit scantly populated,     
metal-poor tail and a main component peaking at [Fe/H] $\sim$ -0.6.   
The most noteworthy characteristic of the abundance distributions is their   
overall similarity in all the sampled fields, covering a wide range of   
environments and galactocentric distances. Nevertheless, a few interesting   
differences and trends emerge from the general uniformity of the   
metallicity distributions. For example, the median $[Fe/H]$ shows a slight   
decrease with  distance along the minor axis (Y) up to $Y\simeq 20 \arcmin$,    
but the metallicity gradient completely disappears beyond this limit.   
Also, in some fields  a very metal-rich ([Fe/H] $\ge$ -0.2) component    
is clearly present.   
    
Whereas the fraction of metal-poor stars seems to be approximately constant      
(within few percent) in all fields, the fraction of metal-rich and,    
especially, very-metal-rich stars varies with position and seems to be     
more prominent in those fields  
superposed on the disk and/or with the presence of streams or substructures  
\cite[e.g.][]{iba01}. This might indicate and possibly trace interaction   
effects with some companion, e.g. M32.   
   
}    

\titlerunning{The Andromeda Project I}
\authorrunning{Bellazzini et al.}

\maketitle 
    
\section{Introduction}    
    
As the nearest bright spiral and the most prominent member of the Local Group,    
M31 has played a central role in our evolving understanding of stellar    
populations. Most notably, Baade's (1944) identification of Population II    
in M31 initiated a true paradigm  shift  which in large part remains    
valid to the present day.    
  
Therefore, studying in detail the stellar populations in M31 and comparing   
their properties to those of their Galactic counterparts is obviously very   
important to understand the formation and evolution history of these two  
galaxies. Our present knowledge already points to some interesting   
differences, supported by observational evidence \citep[see]  
[and references therein]{hh02,fer02}, e.g.:    
  
i) The globular cluster population in M31 is $\sim$ 3 times larger than in    
the Milky Way, and on average more metal rich   
\citep{mou86,du94,du01,barmby,per02};   
  
ii) Also the abundance of the field population, in those few cases where   
it is derived from calibrated colors, seems generally higher than in the   
Milky Way, even at large galactocentric distances where the halo population   
dominates \citep[cf.][and references  therein]{du01, fer02, rich96}.   
A similar result was found  for the halo stars in NGC 5128, with a striking   
difference compared to the Milky Way \citep{hh00,hh02}. This has an impact on   
galaxy formation models, in particular those based on accretion of disrupted    
satellites, since dwarf galaxies are metal poor    
\citep[][and references therein]{dco02}.    
   
iii) \citet{pri94} found that ``a  single de Vaucouleurs  law  luminosity   
profile can fit the spheroid of M31 from the inner bulge  all the way out   
to the halo''. However, clear evidence of ``disturbances'' such as spatial   
density and metallicity variations has been found, that can be interpreted   
as streams and remnants of tidal interactions  \citep{iba01, fer02}.    
   
The available data, therefore, would point toward interaction/merger    
events as non negligible factors in the building of the M31 halo.   
Therefore, very deep and detailed studies of the stellar  populations and    
abundance distributions in the M31 halo and outer disk are needed,   
as they can play a fundamental r\^ole in explaining the formation of the   
M31 halo.

The advent of new technology detectors and telescopes, in particular the Hubble   
Space Telescope ($HST$), has boosted a new generation of studies, providing   
a much deeper insight in the understanding of the stellar content and   
evolutionary  history of this galaxy.    
In line with these recent efforts    
\cite[e.g., see][and references therein]{du94,mo94,cou95,rich96,hof96,ho98,   
guha00,fj01,du01,ste,svd,fer02,wil02},    
we present here the first results of a large   
systematic survey of the stellar population in the disk and halo of M31   
performed with the Wide Field and Planetary Camera-2 (WFPC2) on board $HST$.   
   
The $HST$ program GO-6671  (PI R. M. Rich) was aimed primarily at imaging      
with the WFPC2    
a sample of bright globular clusters in M31 spanning a range in     
metallicity.  The globular cluster target is placed on the Planetary   
Camera (PC) field, while the adjacent halo or disk field population of M31   
falls on the 3  Wide Field Camera (WFC) chips.   
Adding to these  7 more fields adjacent to globular clusters, whose images      
are taken from the $HST$ archive, brings a total of 16 deep fields imaged by     
$HST$, over a distance of about 4.5 to 35 kpc from the nucleus. More M31    
fields are available in the $HST$-archive, but we analyze here only    
those obtained in a strictly homogeneous way (i.e. same camera    
and filters, similar exposures).     
    
Photometry of the clusters is considered in a separate paper     
\citep{rich03}. In this paper we consider only    
photometry of the fields imaged by the WFC (Wide Field Camera) chips.       
Our photometry generally reaches to $\sim 1 $ mag fainter than the      
horizontal branch.    
    
Our primary aim is to derive the abundance distribution of the M31 fields     
assuming that the population is old and globular cluster-like.      
This assumption is well justified by deep photometry \cite[e.g.][]{rich96}     
that shows the subgiant luminosity function to approximately match that of     
the old  globular cluster 47 Tuc.      
The abundance distribution of the M31 field stellar population is then     
derived by interpolating between empirical    
template globular cluster RGB ridge lines,     
a technique first applied by \citet{mou86}  and later used in most     
of the subsequent studies, though with significant differences.

The paper is organized as follows.  Section 2 discusses the   
properties of our field locations within the Andromeda galaxy.    
Section 3 discusses our observations and data analysis.    
Section 4 presents the individual color-magnitude   
diagrams, including an analysis of the young main sequence (blue   
plume) stellar population.  Section 5 describes our method   
for deriving the abundance distribution of the old stellar   
population.   
Section 6 reports the abundance distributions,   
while Section 7 considers the error   
budget (sensitivity to parameters such   
as reddening, distance modulus and abundance scale).   
We compare our abundance distributions with those in the   
literature in Section 8, while Section 9 considers how the   
detailed abundance distributions vary with position and environment.   
We summarize our results in Section 10.   
    
\section{The sample}    
    
M31 is a huge object in the sky, its optical angular diameter as reported by    
\citet{ho92} is $240 ~arcmin$, the absolute magnitude $\sim M_V=-21.0$. The     
galaxy has an inclination angle of $i=12.5 \deg$ \citep{sim78,pri94}, and the    
apparent axis ratio is $\sim 0.65$ \citep{wk88,dev58}.    
 Three main stellar structures have been identified: a wide exponential disk    
with spiral structures, a bright central bulge and an extended halo.    
    
The peculiar inclination has prevented any firm conclusion on the presence of    
a thick disk as the one observed in our own Galaxy     
\cite[but see][for a different view]{svd}.      
Following \citet{pri94} we will consider the bulge and the halo as a     
single component, the {\em spheroid}. We stress that    
this choice is made for the sake of simplicity and doesn't imply any prejudice    
about the formation of the halo and the bulge or the relation between them.    
    
In Fig. 1, the positions of the observed fields are overplotted on a $\sim    
1\deg \times 1\deg$ image of M31, obtained using the Tautenburg Schmidt    
telescope. Our fields sample the disk and halo of M31 over a wide range of     
radii and positions, from $\sim$ 20 arcmin ($\sim$4.5 kpc) as far as     
$\sim$ 150 arcmin ($\sim$ 35 kpc) along the major and minor axes.     
    
    
Table 1 gives the observed fields, which are designated by their associated     
globular cluster (using both \cite{sar77} and \cite{bat87} designations).    
The rectangular coordinates X and Y [arcmin] are reported in column 3 and 4,    
where the X axis coincides with the major axis of M31, and the Y axis coincides     
with the minor axis. $R=\sqrt{X^2+Y^2}$, the radial distance from the     
center of the galaxy, is reported in column 5, and  the adopted reddening    
$E(B-V)$ is reported in column 6 (see sect. 2.2). Finally, we report in     
column 7 the approximate fraction of spheroid stars over the total      
($F_{Sph}$), as estimated in sect. 2.1 (but see also Sect. 9,10 for     
discussion).

\subsection{Expected contributions to the observed samples}    
    
Only the outermost lines of sight in M31 sample a pure halo population,     
and clear traces of a disk population are found even at those locations     
somewhat outside the faint disk isophotes.    
Therefore, we image some combination of disk and halo in most fields.    
    
It is possible to derive an approximate estimate of the relative    
contribution of the disk and the spheroid along a particular direction using    
the models by \citet{wk88} which reproduce very well the surface brightness    
profiles of both components over a very wide range \cite[see also ][]{pri94}.     
The fraction of light must scale as the fraction of light contributors     
(stars), at  least to first order. Since the \citet{wk88} models refer     
to the major axis and minor axis profiles, reasonably good estimates     
can be obtained for the fields that lie in the vicinity of the axes.    
    
The fields G33, G87, G287 and G322 can be considered to lie on the major axis.     
All of these fields are largely dominated by disk population.     
However, while G33, G322 and G287 are sufficiently far from the    
center and suffer only a marginal contamination by halo stars ($\sim 7$\% for    
the first two fields and $\sim 12$ \% for the latter), the growing importance    
of     
the bulge has some influence on the G87 field ($\sim 18$\% of contribution by     
the spheroid).    
    
The fields G76, G119 and G272 are clearly projected onto prominent features of    
the disk and their distance from the major axis is $< 12^\prime$. Given their    
position with respect to the bulge, the contamination by spheroid stars can be    
assumed to be $< 10$\%. Obviously, the model of \citet{wk88} doesn't include    
disk substructures such as spiral arms, and some fluctuation is possible.    
In particular the line of sight of G76 samples the $R=40 ^\prime$ring,    
the site of the most vigorous star formation in M31 \citep{ho92,wil02}.      
    
The fields G108, G64 and G58 are relatively close to the NW arm of the minor    
axis, so the contribution of the various components to their population mix can    
be more correctely estimated from the minor axis profile. The relative    
contributions of the spheroid are $\sim 60$ \%, $\sim 65$ \% and $\sim 70$ \%,     
respectively.     
These lines of sight intersect the halo, but the contribution from the disk     
population is {\em very} important.    
    
For the other fields a clearcut estimate based on the axis-oriented profiles    
cannot be done. However, all of them are farther than one degree from the center    
of M31 and most are expected to be dominated by the halo.     
In particular G319,  not very far from the minor axis and at $Y=-68.9$ arcmin,    
can be considered a ``pure halo'' field;     
G11, G219 and G351 all have $Y>40 ^\prime $,     
and the contamination by disk stars is expected to be low ($< 10$ \%).    
G105 has $Y=-29.8^\prime$ and, judging from the minor axis profile the    
minimum contribution from the disk population should be $\sim 20$ \%.    
Finally, G327 has $Y=5.1^\prime$ but is more than $2 \deg$ from the center of     
the galaxy. The halo contribution as formally estimated from the major axis     
profile (with considerable uncertainty) is $\sim 6$ \%, but the large distance   
from the center of the galaxy suggests that it may be considerably larger than   
this.    
    
Contamination by foreground stars belonging to our own Galaxy and by     
background distant galaxies and quasars is also possible. This issue has    
been extensively discussed by \citet{hof96} who found the contamination from     
such sources on the final CMDs and luminosity functions to    
be negligible.  We repeated their tests and confirm their results.    
    
\section{Observations and Data Reduction}    
    
The observational material is described in detail in Table 2, which gives:    
(1) the field identification, (2,3) right ascension and    
declination of the associated PC pointing, (4) date of the observations,    
(5,6) identification number of the $HST$ program that obtained the observations,    
and name of the program PI, and (7) filter and exposure time of the    
available CCD frames.    
    
Most of the images are from the WF chips of exposures in which the PC was     
centered on the globular cluster (which was the prime target).      
One globular cluster, G327, was missed; our field is 36$^\prime$    
from the cluster but we preserve our usual naming convention.     
    
The G272 and G351 fields are parallel WFPC2 images from GO-5420,    
which imaged G280 and G351 using the FOC.  Both fields are located     
at $<5^\prime$ of the respective clusters.     
    
The G287 field was accidentally observed twice: G287b is slightly    
rotated relative to G287a.  These data enable us to test our    
photometry (Sec. 3.4).  Although we consider 17 fields, our    
survey actually has only 16 independent lines of sight.    
    
The principal characteristic we want to stress here is the great homogeneity     
of the dataset.  Of the 17 fields, 11 were imaged as part of the same    
proposal, and the integration times in each filter are identical.    
The other six image sets have been selected because of similar    
exposure times and choice of filters. Furthermore, all of the repeated     
exposures in each set were taken at fixed pointing and we verified that     
the relative position of each image within a data set coincides with the    
others to within a small fraction of a pixel.    
This allowed us to produce stacked images without any shift and/or    
interpolation (see below).    
    
The distribution of stars within each field is very homogeneous over the    
scales sampled by the WFPC2 camera.  Image crowding is an issue in all cases,    
with significant variations between the different fields, ranging from  640    
resolved sources/sq.~arcmin in the outermost halo fields, to 13000 resolved    
sources/sq.~arcmin for those fields nearest the nucleus. Obtaining good    
photometry in such conditions is very difficult, since even the best    
PSF-fitting packages are subject to significant errors both in the photometry    
and in the interpretation of the detected sources. This class of problems are    
the most subtle and dangerous because they permit contamination by large    
numbers of spurious sources \cite[e.g. see ][]{ja99}.     
Since most of our fields are not lacking in sample size, we imposed   
quite severe selection criteria in order to clean our sample from    
spurious or badly measured sources. In some cases, up    
to 20\% of the detected sources are rejected,  under well defined    
and fully reproducible criteria which are described in the following    
subsections.    
    
After several tests that allowed us to optimize all the parameters    
involved in any step of the data reduction process, we assembled a fully     
automated pipeline where the only input data are an estimate of the background    
level of the images, the exposure times and the dates of the observations. The    
pipeline carries out all the steps that are described in detail in the next    
subsections, from the PSF-fitting photometry on the stacked and cosmic ray    
cleaned images, to the production of the final calibrated and selected     
catalogue, every step being carefully checked.     
    
The only key step requiring manual intervention is    
the determination of aperture corrections, where obvious bad    
measures or erroneous associations must be removed by hand (see sect. 3.2).    
Thus, also the data reduction process is strictly homogeneous for every image    
set. Its reproducibility has ben carefully tested by comparing the results    
obtained by different people reducing the same image set, and by the comparison    
of the results of the independent reduction of the two overlapping fields G287a    
and G287b. We stress the homogeneity of the data set and of the data    
analysis procedure because we believe that this is a fundamental issue for a    
systematic study, as the one presented in this paper.       
    
\subsection{Relative photometry}    
    
All the analysis was performed on the bias subtracted and flat field corrected    
frames from the STScI pipeline. The overscan region was trimmed and each set of    
repeated exposure images was coadded and simultaneously cleaned by cosmic ray    
spikes with the standard utilities in the IRAF/STSDAS package.    
    
The PSF-fitting photometry was performed using the DoPHOT package    
\citep{dophot}, running on two Compaq/Alpha stations at the Bologna Observatory.    
We adopted a version of the code with spatially variable PSF and modified by P.    
Montegriffo to read real images. A quadratic polynomial has been adopted to    
model the spatial variations of the PSF. The parameters controlling the PSF    
shape were set in the same way as \citet{ols98}, that successfully applied     
DoPhot to the analysis of WFPC2 images.    
    
Inspection of the F814W images shows that they have a higher S/N in general    
than F555W despite their being exposed for equal times in most cases.    
This is not surprising given that most of the detected stars    
are old, metal rich RGB stars.    
A few preliminary tests convinced us to use the F814W frames for the source    
catalog, classifying as valid those sources brighter than 3 times     
the local background noise.    
We then forced the code to fit the same sources in the     
F555W images, using the so called ``warmstart'' option of DoPHOT.     
    
Since DoPHOT provides a classification of the sources, we retained only the    
sources classified as stars (type 1, 3 and 7). We cross-correlated the F555W     
and F814W output catalogues with a tolerance of 1 pixel and finally     
produced a catalogue containing, for every image set, the positions in the     
frame, F555W and F814W instrumental magnitudes and the associated errors, as     
well as a parameter connected with the shape of the    
sources \cite[{\em ext = extendedness parameter}, see ][]{dophot}.    
We found that in all cases the bulk of the sources classified as stars are     
confined in the range $-20.0\le ext \le 20.0$, and consequently we excluded     
from the catalogue all the outliers.    
    
The CTE corrections have been applied according to \citet{whit99}.    
    
\subsection{Calibrations}    
    
To transform the instrumental magnitude from the PSF-fitting photometry to the     
Johnson-Cousin and/or STMAG system it is necessary to apply the aperture     
corrections to a radius of $0.5$ arcsec, i.e 5 WF pixels \citep{holtz95}.    
    
Aperture photometry at a radius of 5 px was obtained for all the    
stars whose peak intensity was higher than ten times the local background    
noise and having no detected companion within a 15 px radius,    
using the Sextractor package \citep{sex}. Average aperture corrections were   
obtained for each field using these stars.    
Deriving reliable aperture corrections in such extreme (and uniform) crowding    
conditions is very difficult [see, for instance, \citet{ja99}].    
Furthermore, because of the heavy undersampling of the PSF in the WF cameras,   
the flux sampled in the 5px aperture is strongly dependent on the exact     
position of the peak of the PSF within the central pixel of the star    
image \cite[see][]{bi96}. Thus there is no unique correction for all the stars    
but one  has to rely on an average correction.   
   
The total uncertainty in the absolute photometry introduced by the aperture    
corrections is in general $\pm 0.05$ mag, but can reach $\pm 0.11$ mag in the    
worst cases, a non negligible uncertainty indeed. However, these are the     
intrinsic and unavoidable limits of the WFPC2 as a photometer, at least in the    
presence of heavy, {\em uniform} crowding.    
It is interesting to observe that all the independently derived corrections     
are very similar (i.e. standard deviations are rather small), a fact that     
testifies the stability of the whole aperture corrections procedure.    
No trend of aperture corrections with position in the frame was detected to    
within the quoted uncertainties.     
    
Once the aperture corrections were applied,  the instrumental magnitudes    
have been reported to the Johnson-Cousin and STMAG systems using the standard    
calibration relations provided by \citet{holtz95}.    
    
\subsection{Cleaning from contamination}    
    
In Fig. 2 the errors in the relative photometry for all the measured sources    
are shown as a function of magnitude. It can be readily appreciated that the    
mean uncertainties and the limiting magnitudes are determined primarily by    
the crowding conditions, and differences in exposure time are much less    
important. For all cases errors are smaller than $0.1$ mag for $V<24.5$.     
A check of the reliability of the errors    
provided by DoPHOT will be described in the next subsection.    
    
    
All the stars with an associated error (either in V or in I magnitude) larger    
than three times the average error at their magnitude level were excluded     
from the sample, as well as all the stars having $err_V \ge 0.6$ mag and/or     
$err_I \ge 0.6$ mag. The average error as a function of     
magnitude was calculated     
applying a $2-\sigma$ clipping average algorithm on 0.5 mag wide boxes.     
The lines superimposed on the plots of Fig. 2 represent the $3\times err_V$     
or $3\times err_I$ thresholds actually adopted.     
Errors much larger than the mean can    
originate from a number of reasons: bad interpretation by the PSF-fitting    
algorithm, exceptionally high and/or variable background, proximity to a very    
bright and/or heavily saturated source, proximity to chip defects, partial     
saturation, etc. In any case, our aim is to prevent contamination by spurious    
sources and this selection criteria is a very effective one.    
    
A careful comparison between the derived catalogues and the corresponding     
images showed us that there were still three categories of undesirable    
sources that passed all the selection criteria applied, i.e.:    
    
\begin{enumerate}    
    
\item Spurious stars detected in diffraction spikes and coronae around     
heavily saturated stars (type a).    
    
\item Faint spurious stars detected in     
luminous extended background galaxies (type b).      
    
\item Faint background galaxies misinterpreted by the code as stars (type    
c).    
    
\end{enumerate}     
    
Identification and hand removal of the type a and b sources was completed for    
all the fields with $n_{source} \le 35000$ stars (see Fig. 2).     
In the same fields    
some type c sources have been indentified by accurate visual inspection of the    
frames, but some faint galaxies probably remain in the final catalogues.    
For the remaining fields ($n_{source} \ge 48000$) the crowding    
prevented any further cleaning. However the spurious sources represent a    
negligible fraction of the samples: 3545 sources of type a,b and c were    
removed over a total of more than 100000 in the inspected fields ($< 3.5$ \%).    
    
    
Fig. 3 shows the rejected sources superposed on the CMD of the    
G64 field. Most of them are type a sources that lie upon the real stars    
distribution, mainly populating the region $(V-I) \le 1.5$, while sources of    
type b and c are redder than this value. The blue region of the diagram is    
little affected by this kind of contamination. If the faintest objects are    
excluded (those with $I\ge 26$), only $10$ \% of the spurious sources    
populate the $(V-I)<0.8$, $65$ \% are found in the range $0.8\le (V-I)\ge 1.5$    
and $25$ \% are redder than $(V-I)=1.5$. Most of the spurious sources are    
rather faint, $58$ \% of them lie below the red HB clump ($I<24.5$), while the    
upper RGB is nearly free of contamination, only $8$ \% of spurious sources    
being brighter than $I=23.5$.    
    
We can safely conclude that any spurious sources possibly surviving these    
cuts could not affect the interpretation of our CMDs,     
that are largely dominated by bona fide stars; in particular, the RGB and    
red HB that we consider in this study are clean.

\subsection{Reproducibility of measures and reliability of errors: a direct    
test.}    
    
The large overlapping area of the G287a and G287b fields made it possible     
to perform a    
direct test of the reliability and reproducibility of our measures.     
As this is the most crowded field of the whole set, it assures that we are    
testing the entire procedure under the most unfavorable conditions. Thus the    
uncertainties derived from this comparison are conservative.    
    
Both images were independently reduced, from the raw frames to the    
calibrated and selected catalogue, including aperture corrections.    
The comparison was finally performed for each WF chip separately.    
First, it has to be noted that virtually all the stars in the overlapping areas    
were measured in both image sets and independently survived the selections.    
Thus the detection and selection criteria are very stable and reliable.    
    
    
In the upper two panels of Fig. 4 the magnitude differences as a function of the    
G287a magnitudes are shown, for all the stars in common. Note that in 4 out of    
6 cases, the absolute average differences are $<0.02$ mag and in all cases they     
are $<0.05$ mag. This demonstrates that the uncertainties in the field-to-field     
relative photometry are fairly small in the final calibrated catalogue,     
i.e. the whole sample of 17 data sets is tied to a common photometric system     
that is homogeneous to $<0.05$ mag. The reproducibility of the results     
is excellent.    
    
Individual uncertainties in the relative photometry are nearly as important as    
the measurements themselves. DoPHOT, as well as many other popular photometry    
packages, derives the errors in each measurement mainly from the signal to     
noise ratio of the fitted images and from the errors in the fitting procedure.     
While    
the approach is formally correct and usually produces a correct ranking of the    
quality of the measures, it is not guaranteed that the absolute estimate of the    
error is correct. The ideal estimate should be derived from the dispersion of    
many repeated measures, a procedure clearly not viable in the present case.    
In the lower panels of Fig. 4 the differences between the errors from DoPHOT     
($\epsilon_{Va}$, $\epsilon_{Ia}$) and the error on the mean from the two     
measures obtained for the stars in common ($\sigma_{<V>}$, $\sigma_{<I>}$)     
are averaged over 1 mag intervals and plotted versus V and I magnitudes.     
The two quantities are well correlated and    
their difference is generally lower than 0.03 mag. Furthermore, DoPHOT seems to    
slightly overestimate errors, thus the most dangerous occurrence is prevented,     
i.e. drawing conclusions that are not supported by the {\em true} intrinsic    
accuracy of the data.

\section{The Color Magnitude Diagrams}    
    
Final CMDs for all of the 16 fields are shown in Figures 5 and 6.    
The diagrams are ordered (from top to bottom and from left to right)    
according to the projected distance of the field from the major axis X (see    
Fig.1). The only exception is the field G327 that is the outermost one (130     
~arcmin from the center of the galaxy) but is relatively close to    
the major axis and has been plotted as last. Note that nearly the same order     
is obtained ranking the CMDs by increasing contribution from spheroid stars,     
as reported in column 7 of Table 1.       
    
The limiting magnitude depends mainly on crowding. The    
deepest photometry reaches $I\sim 27$ (G351) but on average is about     
$I\sim 26$. Stars brighter than $I\sim 20.5$ are partially saturated.     
Despite the large differences in positions among the fields, the diagrams are    
quite similar if one takes into account the differences in the number of    
sampled stars.    
    
    
\subsection{The main morphological features}     
    
Our color-magnitude diagrams display the following common features:    
\begin{itemize}    
    
\item    
A {\it red and broad RGB}, ranging from $(V-I)\approx 1$ to $(V-I)\approx 4$,     
with the RGB tip near $I \approx 21$, and a prominent {\it red  clump} (RC)     
on the RGB and HB, at $(V-I)\approx 1$ and $I\approx 24.3$.   In all   
fields the upper part of the RGB bends towards redder colors and fainter  
magnitudes   in the $V$ vs $(V-I)$ CMD, a   
characteristic seen in metal rich globular clusters and caused by   
TiO blanketing in metal rich first ascent stars.    
These characteristics are consistent with a relatively high metallicity and a wide abundance range.    
    
The presence of the dominant red clump   
 in {\it all} of the halo fields is {\it prima    
facie} evidence for an old metal-rich stellar population in the halo, which is     
most likely a mixture of intermediate age RGB stars over the full metallicity     
range and metal-rich old halo HB and RGB stars. The high metallicity is also     
consistent with the slope of the  RGB.      
    
\item    
A {\it blue plume} at $(V-I) < 0.5$, reaching in some cases $I\sim 20$,     
is very evident in some of the CMDs. This feature can be identified with the     
main sequence of an intermediate/young population (YMS).     
    
Despite the large number of sampled stars, the YMS is barely noticeable in the     
G287 and G87 fields, while it is clearly present in G33, G322, G119, G272, G108     
and G64, and it is strongly present in the G76 field.      
    
The existence of a sparse YMS cannot be excluded in    
the fields G58, G11 and G351. In the last two cases the low total number of    
stars in the sample prevents any firm conclusion in this sense. In any case,     
the good anti-correlation between the prominence of the YMS and the estimated    
$F_{Sph}$ is remarkable.    
    
\item     
A small number of {\it blue HB stars} at $(V-I)\sim 0.5$ and $I\sim 24.5$ is     
present in many of the halo fields (e.g. G11, G105, G219, G327, G319, G319    
and G351), and could be      
also present, but hidden within the YMS population, in the other fields     
(see e.g. G64).      
    
These stars indicate the presence of some globular-cluster age metal poor     
stars, because presumably only stars older than 12 Gyr can reach the blue HB.    
Their numbers ($\approx 15$\%  of the RHB) are consistent with the fraction    
of metal poor stars in the abundance distributions that we derive from the    
RGB.    
    
\item     
In the CMDs where the YMS is most conspicuous, a {\it red plume} is seen    
between $I\sim 22$ and $I\sim 20$, around $(V-I)\sim 1.5$.     
Because of the strong young stellar population, we interpret stars in this     
region as being red supergiants, the evolved counterpart of the    
YMS stars.     
     
\end{itemize}     
    
We represent the CMD of the G76 field as a contour plot (Hess diagram) in     
Fig. 7, in order  to illustrate more clearly the principal features     
present in all of our fields.     
    
The most prominent feature common to all of the CMDs (other than   
the red giant branch) is the     
Red Clump, which shows an elongated structure with a very sharp peak.     
The few stars brighter than $I\sim 21$, beyond the RGB tip, could either be     
cool RSG and/or bright AGB stars, the latter not necessarily indicative of     
the presence of intermediate age populations, [see \citet{gua97}].    
There is also a real possibility that some of these are simple blends    
of normal RGB stars, since this is expected in extremely crowded    
fields [see \citet{ja99}, in particular their figs. 5a and 5d, and see also     
\citet{alv98}].     
    
An accurate treatment of the blends, requiring a large number of     
artificial star experiments, is important for some issues (e.g. to constrain    
the fraction of intermediate-age stars in the halo)  that are beyond     
the scope of the present paper.     
    
    
Figure 8$a$ shows stars from the halo fields G105,G219, G319 and G327 are    
plotted together in the same (V,V-I) CMD, centered on the   
Horizontal Branch.    
To limit the contamination from spurious measures, we plot only the stars    
with photometric errors less than 0.15 mag in both passbands.    
    
The BHB is clearly visible at $V\sim 25.2$, extending from the red HB clump     
to $(V-I)\sim 0$, as previously found by \citet{hof96}.     
A schematic ridge line derived from the HB of the metal poor globular cluster     
M68 [data from \citet{wal94}] is superimposed on the plot, applying a shift     
of 9.44 mag in V and 0.09 mag  shift in color.     
    
   
Fig. 8$b$ shows the CMDs of the fields G11 and G351,    
plotted as in panel (a). A BHB similar to that shown in panel (a)    
is still present, but in this case the feature is partially contaminated by    
fainter stars looking like the base of the YMS blue plume described above, and    
this is the reason why the CMDs are plotted separately.  Considering     
the distance of G11 and G351 from the disk,  the blue stars might also be     
field blue stragglers.     
    
As G11 and G351 are among our deepest fields, we cannot exclude such    
a blue plume from the other fields either.  Given the complexity of the    
M31 halo, it is possible that intermediate-age populations are distributed    
throughout the halo in various tidal streamers, present in some fields    
and not (or less) in others.  The only secure determination of the nature of     
these faint blue stars will be imaging deep enough to reach the old main     
sequence turnoff at $M_V=+5$.     
However, even from the present data sets we can     
already deduce that subtle and possibly important differences may exist in the    
stellar populations of the outer fields of the  M31 halo.

\subsection{The intermediate-young population}   
   
In those fields with a substantial disk contribution, the presence of a blue   
plume is evidence of star formation within the last 0.5 Gyr.  The   
power of our survey is that we can use  deep imaging over a range of fields,   
enabling us to probe well past the few Myr of star formation history   
revealed by the brightest OB stars and HII regions.   
     
Any attempt to derive the star formation history (SFH) from the CMDs and LFs    
would need a detailed analysis based on the comparison with appropriate     
synthetic evolutionary tracks and taking into account all the observational     
effects [see, e.g. \citet{tosi89}, and  \citet{ale99,gall99,hern99} as examples     
of recent applications].  \citet{wil02} performed a similar type of analysis     
in M31     
using 27 fields imaged by $HST$ along the disk,  in order to determine the     
star formation history that best fits his observations.     
    
This type of detailed study is beyond the scope of the present paper; however,     
it is possible to apply a simple analysis to our data and obtain useful     
information to constrain the relevant timescales and stellar masses     
associated with the prominent YMS of many disk-dominated fields.     
This aim can be easily achieved by comparison with theoretical isochrones,     
and  would also help to set a basic interpretative scheme for the CMDs     
we present.    
    
In order to compare our CMDs to theoretical quantities, however, we need to     
make assumptions on two basic parameters, reddening and distance modulus.      
    
\begin{itemize}    
    
\item{$Reddening -$}     
Interstellar extintion toward M31 can be due to dust screens residing either    
in our Galaxy or internal to M31 itself. While a reliable estimate of the    
average reddening due to the Milky Way can be obtained from    
reddening maps \citep{buh}, estimates of the ``intrinsic'' reddening are    
not available and this is one of the major sources of uncertainty in the    
study of stellar populations in M31 [see the thorough discussion by    
\citet{barmby}]. These latter authors attempted to estimate the total reddening    
toward the M31 globulars, calibrating the relations between integrated colors    
(essentially B-V) and metallicity with Galactic globulars for which the amount    
of reddening is known from independent measures.     
    
We have applied a similar relationship    
to the integrated colors and    
metallicities of the globular clusters associated with the fields that are    
projected onto clear disk structures, i.e. that are more likely affected by    
intrinsic extinction.     
For G76, G87, G119, G272 and G287 we find an average    
$<E(B-V)>=0.14$ and $\sigma_{E(B-V)}=0.03$. We decided to adopt this value     
since the single values were remarkably similar and to avoid the introduction    
of noise due to the uncertainties of the calibration.    
    
For all the remaining fields that are distant from the high surface brightness     
regions of the disk we adopt the Galactic reddening toward M31.      
Following \citet{ho92}, we adopt $E(B-V)=0.11 \pm 0.02$ from \citet{mac};     
we use the \citet{dean} extinction law $E(V-I)=1.34E(B-V)$ and     
$A_I=1.31E(V-I)$.    
    
\item{$Distance -$} We adopt the Cepheid distance modulus of $(m-M)_0=24.43     
\pm 0.06$ ($\sim 770$ kpc) from  \citet{fm90}.    
Other distance estimates may be found in Table 3.1 in \citet{syd00}.    
    
\end{itemize}

\subsubsection{The age range of the YMS stars}     
    
Fig. 9 shows the CMD (in $M_I, (V-I)_0$ for the sum of the     
disk-dominated fields G76, G119, and G322 (154,748 stars).      
The stars above and below the $M_I=-2$ threshold are plotted using    
points of different thickness to allow easier recognition of both the    
densely populated features in the lower part of the CMD (e.g. the HB    
clump) and the sparse bright features (e.g the upper MS and the red plume    
of RSG stars).    
    
    
Six isochrones with solar metallicity and helium abundance $Y=0.28$,     
from the set by \citet{bert94}, are superimposed to the diagrams.     
They correspond to ages of 60, 100, 200 and     
400 Myr (from top to bottom, continuous lines), 1 Gyr (open squares) and 12 Gyr    
(open circles).  Our adopted composition seems the most appropriate for    
intermediate/young populations in the disk of a large spiral galaxy.    
    
Before we consider Fig. 9, recall that the brightest MS stars known in    
M31 reach $M_V \sim -6.2$, but the saturation level in our survey occurs     
at $M_V \sim -4$, which eliminates the brightest main sequence stars from our    
sample.    
    
The isochrones of 60, 100, 200 and 400 Myr fit well the observed    
distribution of young and intermediate age stars,  thus constraining     
the age range associated with the sampled YMS.    
    
The RSG branches of these isochrones group between $(V-I)_0\sim 1$    
and $(V-I)_0\sim 1.6$, providing an excellent fit to the red plume     
described in the previous section (figures 6 and 9).    
The 1 Gyr isochrone illustrates clearly how intermediate age AGB stars can    
populate the region of the CMD immediately above the RGB Tip.    
    
As a final consideration we note that RGB stars as red as (or redder than)   
the 12  Gyr isochrone are clearly present in the composite CMD of Fig. 9.   
The obvious  conclusion is that very old and very metal rich stars are   
found in the disk of  M31.

\subsubsection{A Classification Scheme for the Blue Plume Population}    
    
In order  to model the main sequence luminosity function we have grouped     
stars by luminosity bins so that they crudely sample different mass ranges.    
We can use these counts to explore the relative importance of the    
YMS in different fields.    
    
In Fig. 10 we overplot 3 boxes on  the (V,V-I) CMDs of the G64 field,    
as an example.    
These boxes are rather large and clearly separated in order to include a    
portion of a particular sequence independent of any slight difference in     
reddening or metallicity among the different fields.    
    
    
The regions are defined as follows:    
    
\begin{enumerate}    
    
\item {\bf Yu}: ($23.5 < V \le 22.5$ and $-0.3 \le V-I \le 0.6$)    
samples the upper YMS. The number of stars in this part of the CMDs are    
indicated as $N_{Yu}$.    
    
\item {\bf Yd}: ($24.5 < V \le 23.5$ and $-0.3 \le V-I \le 0.6$)    
samples the lower YMS. The number of stars in this box are    
indicated as $N_{Yd}$.    
    
\item {\bf C}:  ($24.5 < V \le 23.5$ and $0.9 \le V-I \le 2.6$)    
samples the RGB in the same magnitude range as Yd.     
The number of stars in this box are indicated as $N_C$. We will use $N_C$    
to normalize the star counts in the previous boxes.    
     
\end{enumerate}    
    
According to the same set of isochrones shown in Fig. 9 it can be stated that    
the {\bf Yd} box samples stars younger than $\sim 0.5$ Gyr while the {\bf  Yu}    
boundary samples stars younger than $\sim 250$ Myr.     
We thus derive the following indices::    
    
\begin{itemize}    
    
\item $(N_{Yu}+N_{Yd})/N_C$: ranks the fields according to the    
relative importance of the YMS population.    
    
\item $(N_{Yu}/N_{Yd})$: since more massive (i.e. younger) YMS stars are    
expected to reach brighter magnitudes, this ratio tests differences in     
the recent star formation history between the different fields.    
    
\end{itemize}    
      
The errors in the star counts have been estimated according to Poisson    
statistics, and we propagate the errors to derive errors in    
the indices. It is obvious that our detection of the    
bright MS is most significant  for the most populous, disk    
dominated fields. The fields G11, G58, G105, G219, G319, G327 and G351    
have too few young stars to be worthy of further consideration    
in this sense.  As already noted, this doesn't necessarily    
mean that no young stars are present in these fields, since the low total    
stellar density can prevent the detection of short lived stars. However, a    
comparison with the most similar field having a significant YMS population,    
i.e. G64, shows that these fields are intrinsically deficient of ``Yu+Yd''    
stars by a factor $\sim 2-5$ with respect to G64.    
    
Fig. 11 shows $(N_{Yu}+N_{Yd})/N_C$  (panel a) and      
$(N_{Yu}/N_{Yd})$  (panel b) plotted as a function of the distance     
from the center of the galaxy, deprojected to ``face-on'' $Rp_{[arcmin]}$     
\cite[see][and references therein]{ho92}. The deprojection has been    
obtained adopting $i=12.5 \deg$, and it is justified in the present case     
since we are dealing with $(i)$ mostly disk dominated fields in the plane     
of M31, and $(ii)$ the YMS population which is expected to be located in    
the star forming disk.     
    
The prominent feature appearing in panel (a) is the strong peak in the relative    
abundance of stars younger than $\sim 0.5$ Gyr occurring at     
$Rp \sim 60^\prime$, shown mostly by G76 but also by G119 and G322.      
This corresponds to the $R\sim 10$ kpc ring in which neutral hydrogen and     
virtually any tracer of a young population seem to cluster \citep{ho92,syd00}.     
This main structure of the star forming disk is clearly sampled    
by our fields as a very significant enhancement in the recent star formation    
rate: there is a factor $\sim 35$ in the relative abundance of YMS stars    
between the nearly quiescent inner regions (fields G87, G287) and the peak of    
the ring structure (field G76).    
    
    
In panel (c) of Fig. 11, we show the $(N_{Yu}+N_{Yd})/N_C$ index as a     
function of the absolute distance along the major axis ($|X|$), to provide     
a different view of the results presented in panel (a).     
    
In particular, this panel allows a direct comparison with the distribution     
of disk tracers as depicted by \citet{ho79} (see his Fig. 7 and 8).     
The agreement is indeed remarkable, and the general    
picture assembled by \citet{ho92} is confirmed by our survey, at least to the    
extent permitted by our limited spatial sampling\footnote{It is important     
to recall that the work by \citet{ho79} is based on a sample of 403 open     
clusters while we are presently analysing just 9 small fields.}.    
    
Panel (b) shows that also the star formation in the last 250 Myr was     
particularly strong near the position of the G76 field. However, the most    
noteworthy feature is the high $N_{Yu}/N_{Yd}$ value in the innermost    
field G87, indicating that most of the (weak) star formation in this field    
occurred quite recently.     
    
This is particularly interesting in comparison with the otherwise similar     
field G287 that shows a $N_{Yu}/N_{Yd}$ a factor $\sim 5$ lower for the     
same value of $(N_{Yu}+N_{Yd})/N_C$.    
    
As a result of these considerations, we can conclude that the star formation     
rate has varied significantly over the last 0.5 Gyr in different regions of the   
disk of M31.  Although our findings are not based on as sophisticated    
an analysis as that of \citet{wil02} we agree with the conclusions of that   
study.  It is noteworthy that the observable activity in the 10 kpc ring   
corresponds to recent significant star formation, and that recent star   
formation activity is not necessarily correlated with a high total rate   
of star formation.  Surveys with the $HST-ACS$ and more sophisticated   
modelling will greatly improve our understanding of this issue.

\section{Assumptions and procedures for metallicity determinations}    
    
It is well known that the color of the RGB of an old Simple Stellar Population     
[SSP, i.e. an ensemble of stars sharing the same age and chemical composition,    
see \citet{rebuz86}] is mainly affected by the abundance of heavy elements and    
only at a lesser extent by the age of the population. The influence of age    
becomes smaller and smaller with increasing age and it is almost negligible for    
ages in excess of $\sim 10$ Gyr. In principle, the metallicity of a RGB star is    
uniquely determined by its color and luminosity, once its age is known and    
a suitable calibration is available. The obvious application of such principle    
is to derive the metallicity distribution for a given population from the    
distribution in color and magnitude of its RGB stars \cite[see][]{sav00}.     
    
However, when dealing with a Composite Population (CP), i.e. a mix of stars     
of different ages and metallicities, there are two main factors that can     
affect the correct recovery of the underlying metallicity distribution:    
    
\begin{itemize}    
    
\item {\it AGB sequences}, in general, run nearly parallel to the RGB but are    
slightly bluer. In the absence of any useful criterion to exclude them from the    
sample, they would appear as more metal deficient than their parent population,    
introducing a bias in the metallicity distribution \cite[see][]{hof96}.

The existence of a blue HB    
\cite[as well as the results of the spectroscopic survey by][]{guha00}    
tells us that a metal poor    
old population exists for sure. Hence, the blue side of the RGB must contain    
their metal poor giant precursors. If this is the case, contamination from     
AGB stars is necessarily  very small  since their  lifetimes are    
significantly shorter than lifetimes on the RGB.   
In fact, for a solar metallicity      
population of age 15 Gyr, the number of RGB stars is predicted to    
overwhelm the number of AGB stars by a factor $\sim$ 40 \citep{alv98}.    
Based on the above considerations, only a very small fraction of our metal     
poor giants may, in principle, be misclassified AGB stars.     
    
\item {\it Age differences} at fixed metallicity produce a widening of the RGB     
that could be erroneously interpretated as a metallicity range.     
This effect is rather     
subtle, since it depends on the whole star formation and metal enrichment     
histories of the composite population under study.    
    
For example the     
RGB of an old population (say $\sim 12$ Gyr) of solar metallicity can be     
contaminated by stars from a population of the same metal content but several    
Gyr younger that would simulate the presence of older metal poor stars    
[see sec. 3.6.1 of \citet{syd00}].    
    
The metallicity distributions (MDs) derived from the color distributions     
of RGBs in our CMDs are based on the assumption that the observed RGBs are     
dominated by old stars.     
    
This hypothesis is not at odds with observational evidence [see     
\citet{tin71,syd00,ho92}, and references therein], and all the numerous    
attempts made in previous studies are based on it     
\citep{rich96,hof96,pri94,ri90,mo94,ja99,ja00,du94,cou95,du01,fj01,hh00,hh02,      
svd}.      
    
However, our previously stated caveats should be kept in mind when drawing     
any conclusion from the derived abundance distributions.     
This is especially true for the abundance distributions measured in those    
fields dominated by disk stars.      
    
\end{itemize}    
    
In addition to the above issues related to a composite population, there is a     
third problem one must take into account, i.e. the differential reddening     
that might affect the photometric data especially in the disk dominated fields.    
    
Differential reddening does not seem to affect significantly any of our     
fields at a cursory inspection: for example, the wide RGB    
is found in distant halo fields, and no dispersion of the red clump    
along the reddening vector is seen in any field. Further, the blue    
main sequence, where present, is not similarly broadened.    
    
However, we tested possible variations in interstellar/intergalactic extinction     
across    
the observed fields by comparing the CMDs obtained from different subsections    
of each single WFPC2 field. In every case we failed to detect any difference in    
the location of the main branches, and we conclude that variations of the    
extinction across the observed fields, if any, are negligible with respect     
to the intrinsic width of the branches and the observational errors.    
    
Variations of the extinction along the line of sight could also affect the    
CMDs, but this effect would be more subtle.  Dust structures in M31 might be    
embedded in the population, with some stars in front and some behind the cloud.    
However,  the striking similarity between MDs    
(see Sect. 6.2) of fields sampling dense and active    
regions of the disk (f.i. G287, G33, G76) and those sampling outer halo lines    
of sight (f.i. G319, G351, G1), that are expected to be unaffected by M31 dust,   
argues against this type of differential extinction.    
To obtain such a result from a dataset where differential reddening has a    
significant effect in some fields and none in others would imply an implausible    
fine tuning between differential reddening, metallicity distribution and star    
formation history.    
    
We conclude that differential reddening has at most a marginal effect on our    
CMDs and MDs.

\subsection{The grid of Galactic Globular Cluster Templates}     
    
In our analysis we shall adopt as a reference system the \citet{cg97}     
(CG) abundance scale, that is tied to high-resolution spectroscopy     
and offers a better guarantee of accuracy and reliability with respect     
to the \citet{zw84} (ZW) metallicity scale, that was based on     
photometric indices. However, the ZW scale has been and still is widely     
used, so we shall comment our results by comparing the effects of both     
scales (see Sect. 7.1).    
    
For a detailed description of the CG vs. ZW scales we refer to \citet{cg97}.    
Here we just note that the CG scale yields more metal-rich values in the     
interval $-2 < [Fe/H] < -1$ (this effect disappearing progressively as one     
moves towards the  edges of this interval) and more metal-poor outside.      
Also, the metal-rich extension towards solar values is still poorly    
sampled and very uncertain for both scales, with CG yielding slightly     
more metal-poor values than ZW.     
    
As we will discuss below, the choice of the CG or ZW metallicity scale     
has little effect on our main conclusions concerning the overall properties    
of the derived MDs. In fact, the same general conclusions can be drawn by    
a direct comparison of the star color distributions with the grid of the     
RGB ridge lines for Galactic GCs of known metallicity.     
However, the {\it detailed} shape of our abundance distributions and,     
therefore, some {\it specific}  conclusions may indeed be affected by this     
choice (and, more significantly, by the adopted globular cluster grid and     
interpolating procedures).    
    
Finally, we caution the reader that {\it systematic} effects on the     
zero-points of magnitudes and     
colors or of the metallicity scales may actually  have the strongest    
impact on the overall picture.     
    
For the metallicity determinations      
we adopt an approach similar to \citet{hof96}, i.e. we compare the observed    
RGBs with a grid of accurately chosen RGB fiducial ridge lines at various     
metallicities, and then derive a metallicity estimate for each star by    
interpolating from the grid of templates. We apply the technique in the     
$[M_I,(V-I)_0]$ plane since in this plane the behaviour of the RGB sequence     
is less susceptible of  ``curvature'' effects \cite[see][]{sav00}.    
    
As RGB templates we adopt the ridge lines of the galactic globular clusters    
NGC 6341 ($\rm [Fe/H]_{CG}=-2.16$), NGC 6205 ($\rm [Fe/H]_{CG}=-1.39$),     
NGC 5904 ($\rm [Fe/H]_{CG}=-1.11$) and 47 Tuc ($\rm [Fe/H]_{CG}=-0.71$)     
from \citet{sav00}.     
    
As noted by all the authors adopting the present approach, the extension     
of the grid to more metal-rich values than 47 Tuc is quite difficult (for     
the lack of suitable candidates) and extremely uncertain (for intrinsic     
uncertainties in the high-Z regime).     
    
After a careful revision of the available data, we choose to adopt    
for the very metal-rich regime two clusters, NGC 6553 and NGC 6528, for    
which sufficiently good V,I data and metallicity estimates    
(for both CG and ZW scales) are available.    
    
Therefore we completed the reference grid with the ridge line    
of:     
    
NGC 6553 ($\rm [Fe/H]_{CG}=-0.16$, \citet{coh99}) obtained     
from the photometric data of \citet{gua97}, and     
    
NGC 6528 ($\rm [Fe/H]_{CG}=+0.07$,  \citet{ccgb02}) obtained     
from the photometric data of \citet{ort95}.     
     
Reddenings and distance moduli are  taken from the    
compilation by \citet{fer99}, since their approach in the measure of    
distance moduli is homogeneous for all the clusters and independent of their    
HB morphology or of the presence of RR Lyrae variables.    
Table 3 includes these values and the metallicities of our   
calibrating clusters.

    
Before proceeding further, we wish to add a few considerations    
on the problems and implications the choice of a different grid may    
have on the subsequent analysis.    
    
All previous studies dealing with a similar procedure of deriving the MDs     
from the color distributions of the giant branches in the CMDs have adopted     
their own custom-tailored recipe and reference  grid (see Sect. 8).    
Some authors have adopted purely  {\it empirical} grids or,    
alternatively, purely {\it theoretical} ones,      
others have adopted  a  {\it mixture} of both, using GC data to set     
the zero-point of a given set of theoretical models, or adding     
suitably calibrated isochrones to empirical templates to cover missing     
parts of the metallicity range.     
 Furthermore, within each adopted procedure, the choice of selected GCs     
and/or set of models differs from study to study.    
 Though the bulk of the main results is probably independent of    
these differences, there is no doubt that the  detailed  shapes and    
properties of the MDs are strongly affected and,    
especially in the very high-metallicity regime, these choices may    
dominate the results.      
    
The reader should be aware that the same original data, transformed into     
MDs with different recipes, may yield significantly different results     
{\it in some details}. Coupled with the intrinsic, quite high uncertainties     
still affecting the observational data-bases themselves, one has to admit     
that all our analyses are still at a rather preliminary stage and far from     
offering an unambiguous detailed comprehension  of this issue    
\cite[see][for discussion]{hh02}.       
    
We have emphasized above that our reference grid relies only on {\it     
empirical} templates, chosen to be a homogeneous and reliable set, and     
sufficiently sampled to cover the relevant metallicity range in     
suitably fine steps.  The accuracy of this approach depends     
only on observable quantities, i.e. photometric and spectroscopic data    
and reddening estimates, whose errors can be known and minimized to some     
extent, within the present possibilities.     
The use of theoretical models, that would be in principle easier     
and more precise, suffers of its own set of problems.     
In fact, while great progress has been made in producing synthetic     
RGB sequences, the physics of late-type stellar atmospheres (e.g.     
convection, alpha enhancement, etc.) and the color-temperature calibration,    
bolometric corrections, etc. are still poorly known, and the use of     
theoretical RGBs might introduce additional uncertainties which are     
difficult to correctly estimate and quantify.

\subsection{The procedure of metallicity determination}    
    
Both the M31 CMDs and the templates are transformed to the absolute plane,    
using the values of reddening and distance quoted in Sect. 4.2 for M31,     
and listed in Table 4 for the templates.     
    
The estimates of metallicity are performed     
on stars having  $-3.9<M_I<-2.0$ and $0.90 < (V-I)_0 < 4.0$. The reasons   
of this choice can be summarized as follows:    
\begin{itemize}    
    
\item{} Lower Luminosity limit: $M_I<-2.0$     
    
to retain the region of the RGB with the highest sensitivity    
to metallicity variations and to avoid contamination by RC stars. This choice    
also avoids the inclusion in the sample of the AGB clump stars, which are     
predicted (and observed) to lie around $-1 < M_I < -1.6$ \cite[see][]{fer99}.    
Thus the more densely populated feature of the AGB is excluded and we can be    
confident that only a marginal fraction of old AGB stars may contaminate our    
metallicity distributions.      
    
\item{} Upper Luminosity  limit: $M_I>-3.9$    
   
to avoid the inclusion of bright AGB stars.    
To examine in detail the possible impact of different bright cuts    
we have carried out several tests (cutting at different     
luminosity thresholds and using different interpolating schemes; the one    
actually used is illustrated in  Fig. 12 and 13).   
These tests have produced insignificant deformations in the MDs, and   
varying the upper luminosity cut has a negligible effect on the basic   
morphology.  
Therefore, in the following we     
shall use the above bright limit because it adds statistical       
significance, especially in the poorly populated halo fields.    
   
\item{} Blue color limit : $(V-I)_0 > 0.90$    
   
to limit the contamination by AGB stars and  young stars (see Fig. 9).    
    
\item{} Red color limit: $(V-I)_0 < 4.0$    
   
to avoid contamination by foreground  and/or background sources.      
    
\end{itemize}

The interpolation procedure has been accurately checked and     
tested,  and proven to perform very well. We estimate that the uncertainty     
in a single metallicity measure is $\pm 0.2$ dex (random error).    
     
Our interpolation scheme is strictly self-consistent only over the metallicity    
range defined by our grid of GGC templates, but we decided to allow a modest    
linear extrapolation to stars slightly bluer than the ridge line of NGC 6341     
and slightly redder than the ridge line of NGC 6528. The allowed extrapolation     
is of the order of the assumed uncertainty in the metallicity estimates,     
i.e. 0.2 dex: stars beyond these extrapolated limits are excluded from    
the final metallicity distribution (MD). The fraction of stars excluded    
from each sample  is in general less  than $ 2-3$ \% (see Table 4),    
and does not have significant effects on  the description of the MDs.     
    

\section{Results: the Metallicity Distributions (MDs)}

\subsection{A first cursory inspection}    
    
We show in Fig. 12 the $[M_I,(V-I)_0]$ CMDs of the upper RGBs for four    
fields, taken as representative of the whole sample: G87 (the innermost one),    
G119 (a disk dominated field with a rich YMS population), G64 (an    
intermediate field with significant contributions from both disk and spheroidal    
components), and G11 (a typical halo dominated field).    
    
The template ridge lines are superimposed to each plot, from left to right:    
NGC 6341 (M92), NGC 6205 (M13), NGC 5904 (M5), NGC 104 (47 Tuc), NGC 6553,     
NGC 6528. The inner frame drawn on each plot shows the region of the     
CMD selected for the metallicity estimate, as described in the previous     
section.     
    
A cursory inspection of Fig. 12 shows the following qualitative     
characteristics:    
    
\begin{itemize}    
    
\item The distribution of the stars with respect to the ridge lines is     
quite similar for the four cases, i.e. the bulk of them     
is bracketed by the ridge lines of 47 Tuc and NGC~6553,   
with a very wide spread.     
Taken at face value, this suggests that the metallicity distributions must     
peak somewhere in the range $-0.7\le [Fe/H]\le -0.4$,     
independently of which metallicity scale is used for the detailed     
abundance estimates. This is in agreement with many previous results     
\citep{mou86,ri90,du94,mo94,cou95,hof96,du01,fer02,hh00,hh02}.     
    
\item  The width of the RGB distribution in M31 largely exceeds the     
observational errors, which are less than 0.10  mag in each bandpass.      
This result too has been found in all previous M31 studies.      
Differential reddening, which could cause such a spread, can be ruled out     
as it was discussed in Sect. 5. Therefore, some    
metallicity (and possibly age) spread must be present.     
    
\item     
The statistical weight of the metallicity distributions varies from field to     
field depending on the density of the stellar population: sparse halo fields     
can suffer from small number fluctuations but are cleaner from contaminants,     
rich disk dominated fields have a larger number of stars but are more prone     
to contamination by younger populations.    
Note however that the adopted selection in color is quite effective in    
preserving the samples from the pollution by undesired stars in such     
fields (see the CMD of G76).    
    
\end{itemize}    
    
For a more quantitative and detailed analysis, we need to investigate the     
metallicity distributions in all our fields.

\subsection{MDs of the individual Fields}    
    
Based on the ridge lines, the distance moduli and the reddenings     
of the template clusters listed in Table 3, and on the     
assumptions we have made about the distance modulus and reddening of M31     
(Sect. 4.2), we have derived the metallicity distributions for all    
the individual fields shown in figures 14 and 15 in the same     
order as in figures 5 and 6.     
    
    
In each panel is reported:     
the name of the field,     
the number of stars used to derive the MD,     
the average metallicity ($\rm [Fe/H]_{ave}$)     
together with the associated standard deviation,     
the median metallicity     
together with the associated semi-interquartile interval, and     
the fractions of stars with     
$\rm [Fe/H]<-1.4$ [Metal-Poor or Young -- hereafter: MPorY],    
$\rm -1.4<[Fe/H]<-0.2$ [Metal-Rich -- MR], and     
$\rm [Fe/H]>-0.2$ [Very Metal-Rich -- VMR],     
respectively (see below).    
      
The most striking property of the MDs shown in figures 14 and 15 is their    
overall similarity. To describe them in a quantitave way (without "forcing"    
the analysis beyond a certain level of speculation), after several    
different tests we decided that the best solution would be to    
schematically subdivide the histograms into three main intervals as indicated    
by the vertical lines drawn in the plots presented in  Fig. 14 and 15:    
    
\begin{itemize}    
    
\item $\rm [Fe/H]<-1.4$, the {\it Metal-Poor or Young group --} MPorY    
    
\item $\rm -1.4<[Fe/H]<-0.2$, the  {\it Metal-Rich group --} MR    
    
\item $\rm [Fe/H]>-0.2$, the {\it Very Metal-Rich group --} VMR    
    
\end{itemize}    
   
The values of --1.4 and --0.2 were chosen because they mark either a    
discontinuity in the MDs (at --1.4), or a discontinuous behaviour    
of the metal-rich tail (at --0.2).     
    
The long "thin" tail reaching metallicities as low as $\rm [Fe/H]<-2.5$    
could be interpreted as due to: (a) very metal-poor old stars     
truly representative of the old halo of M31, though at a trace level,    
and/or (b) young blue stars which mimic a metal poor population,   
especially in the inner  disk fields. This is the reason why we    
have called this group MPorY.    
    
On the opposite extreme of the distributions, in the VMR-regime, at    
$\rm [Fe/H]=-0.2$, especially in some fields (e.g. G287, G87, G33,     
G322, G76, G219, G319), one can see a second discontinuity    
(variable in size from field-to-field) which could be ascribed to the    
existence of a very metal-rich component, added to    
the metal-rich tail of the bulk population.    
    
In a few fields (i.e. G108, G58, G351, G219, G327) one might    
also see another discontinuity at $\rm [Fe/H]\sim -0.8$. However,    
taking into account the quoted intrinsic uncertainties and    
the possible effects induced by the binning size, we are    
inclined to adopt just two cuts. We will discuss in a specific    
section (sect. 9.3) the results of dividing the total sample   
in two sub-samples only, cutting at $\rm [Fe/H]\sim -0.8$.     
    
Before summarizing it may be useful to note that we do not     
report the results of any multi-gaussian  fitting to the data, as    
we found that the degree of discretionality in setting the several    
parameters (number of components, metallicity limits, widths, etc.)    
is rather high compared to the intrinsic quality of the available     
data.     
    
In synthesis all the fields:    
    
\begin{itemize}    
    
\item     
have the bulk of the population located    
within the central metallicity interval, i.e. $\rm -1.4<[Fe/H]<-0.2$, with    
an obvious main peak around $[Fe/H]\simeq -0.6\pm0.20$.     
 Most of the fields    
in Fig. 15 are so distant from the disk as to be true halo fields, yet    
the metal rich peak is the most evident feature of the distribution,    
and the fraction of stars with $\rm [Fe/H]<-0.8$ ranges from only 0.1    
to a maximum of 0.4 (see Table 5).   
To test further this point, we reduced the frames used by \citet{rich96}     
and derived the calibrated photometry of the field of the remote cluster G1     
($R_{arcmin}\sim 150$), with the same pipeline adopted for all fields     
presented here. The resulting CMD  is virtually identical    
to that presented by \citet{rich96}. In Fig. 16 it is shown that the MD of this   
extreme region of M31 is very similar to that of the other inner fields. In   
particular, the mean metallicity is still as high as $\rm [Fe/H]=-0.8$.   
     
\item     
show a long, poorly populated tail spanning the metallicity    
interval $\rm -2.5<[Fe/H]<-1.4$. As one can see in Fig. 8, stars bluer     
(more metal-poor) than the ridge line of G11 ([Fe/H] $\sim$ -1.9) do     
exist and have been detected in all fields at the level of a few percent     
of the total sample. These stars       
are probably representative of the very metal-poor old halo of M31    
in the outer, less disk contaminated fields. In the inner    
disk-dominated regions this subsample most probably includes blue young stars    
which in our procedure (based on colors) may mimic metal-poor objects.    
    
\item     
show a significant population of stars  with $\rm [Fe/H]>-0.2$,     
varying with distance along the Y-axis from $\sim 40$\%  in the most     
central fields to less than  $\sim 4$\% in the most distant ones,    
with a few exceptions (e.g. G219).    
    
\item     
the field of G219 stands out in that its VMR pupolation is stronger     
than might be expected from its location, and the distribution looks     
qualitatively different from the others.     
As discussed  later and shown in Fig. 20 , this field is well superposed    
on the metal rich  tidal tail found by \citet{iba01} and \citet{fer02}    
(their Fig. 7).      
It is interesting to note that the abundance range of this field is similar to     
those located at large galactocentric distance,  only the prominence of the     
VMR peak is greater.    
    
A similar behaviour, albeit with a much wider MD and a less pronounced peak,     
might be present in G319 located not far from the quoted stream (see Fig. 20).

\end{itemize}    
    
Table 4 reports the populations of each metallicity  bin (step 0.2 dex)    
for each field to make available to the reader the grid of the MDs    
for further analysis.

\subsubsection{Is our Very Metal-Poor population real?}      
    
As mentioned earlier, metal poor stars are expected to be present,     
based on the existence of a blue HB.    
If the metal poor tail were mostly due to contaminating younger    
stars it would be expected to reach its minimum extent in the most     
external halo fields where no such stars are expected to be present.    
However,  the fraction of blue objects seems to be fairly constant     
($\sim 2-5$\%), so  we are inclined to interpret them as truly old     
and very metal-poor stars.      
    
Independent confirmation of the existence of metal-poor halo giants      
has been provided by the recent work of \citet{rg02}     
based on Keck/LRIS spectra.     
>From a preliminary sample (obtained from a $16 \times 16$ arcmin$^2$ field     
located at $R=1.6 \deg$ on the M31 minor axis), they selected 24     
{\em bona fide}  M31 RGB stars on the basis of their position in the     
CMD and their radial  velocity, and derived metallicity estimates from     
the strength of the Ca II lines.     
21 of the selected RGB stars have $[Fe/H] \le -1$, 17 of them have     
$[Fe/H] \le -1.5$ and some reach metallicity of $[Fe/H]\sim -2.5$ or lower.    
The result is independent of the assumed metallicity scale. So metal     
poor stars are indeed present in the M31 halo and are not rare.     
    
    
\subsubsection{Is our Very Metal-Rich population real?}      
    
As already noted by simple visual inspection of the CMDs, it is quite     
evident that, at any level of magnitude, a good number of stars    
redder than the ridge line of NGC 6553 ($[Fe/H]=-0.16$) is    
present almost in all fields, and varies from field-to-field      
depending mostly on galactocentric position.     
    
These objects cannot be accounted for by photometric blends or    
by any other conceivable photometric effect. There is also    
no indication to suspect that they are not members    
of the M31 system. If this is true, this population of    
stars does exist in these fields, and should have a very    
high metal content given its location in the observed CMD.    
    
However, due to the combined effects in the (I,V-I) plane of factors like    
e.g. (a) the V,I limiting magnitudes, (b) the saturation for the brightest     
stars, (c) the increasing bending and separation, and the non-monotonic     
behaviour of the ridge lines with increasing metallicity,     
(d) the possible existence of patchy differential reddening, and (e)     
the uncertainties in the adopted metallicity scale that are larger at the   
high metallicity end,     
it is very difficult to assign a precise value of metallicity to these     
objects.     
Both the value of their absolute average metallicity and, especially,    
the detailed distribution over smaller metallicity bins are very uncertain    
and need a much better and deeper analysis,  probably  not feasible    
via photometric means.       
    
\citet{hh99} and \citet{hh00,hh02}  find    
the same result for the halo of NGC 5128.  Interestingly, both the    
metal poor and metal rich peaks they find in NGC 5128 are identical to those    
in the halo of M31.    
    
We emphasize again that any abundance estimate depends on the choice of     
the calibrating templates.  The suggestion of sub-structures     
and a fit to the Simple Model of chemical evolution discussed    
by \citet{hh00,hh02}  are tantalizing, but we caution that     
the present abundances are not sufficiently accurate to allow     
firm astrophysical conclusions with such detail.    
    
\section{Effect of varying assumptions on the MDs}    
    
Metallicity determinations     
obtained via purely photometric data (i.e. stellar/population colors) are     
strongly dependent on various assumptions and uncertainties in fundamental    
issues such as (for known/assumed population age) the choice of the     
metallicity scale, the reddening and the distance modulus.     
    
We explored these effects by re-deriving the MD of a test field (G64) under     
a set of different     
assumptions for the metallicity scale, the distance modulus and the reddening.    
The results of this test are shown in Fig. 17.    
It may be useful, however, to discuss in some detail the possible impact    
of each parameter, separately.    
    

\subsection{The metallicity scale}     
    
Panels (a) and (b) in Fig. 17 show the MD derived for G64 using our standard     
assumptions, and the CG  and  ZW metallicity  scales, respectively.     
Our major conclusions are unaffected    
if we simply shift by about --0.2 dex (i.e. the offset between the two     
scales at mid-metallicity range)  the adopted metallicity boundaries       
(from -1.4 to -1.6, and from -0.2 to -0.4) when passing from CG to ZW.

In particular, we see that both sets of MDs show a    
poorly populated tail reaching as far as $[Fe/H]<-2$.                         
The use of the ZW scale does not affect the median of the MD,     
which remains at [Fe/H] $\sim -0.6\pm0.1$, but it shifts the    
average by about $-0.13$ dex. A    
possible additional peak at $[Fe/H]\sim -1.5$ dex seems to emerge.   
The evidence for the very metal-rich (VMR)    
component remains, though shifted by one bin. In fact, a large    
fraction of the VMR objects in the ZW-scale (see figures 18, 19)    
are found within the interval $-0.4<[Fe/H]<-0.2$  due to the    
non-linear relationship between the two metallicity scales.     
    
Given the importance of this issue, we have then derived the MDs for all     
fields using the ZW scale, to ease the comparison with the previous studies.     
Figures 18 and 19 show the resulting distributions for all the considered     
fields. We note that:      
    
\begin{itemize}    
    
\item     
the bulk of the population is metal-rich and peaks at $[Fe/H]\sim -0.6$, as     
with CG;    
    
\item     
the young/metal-poor population (MPorY) is well detectable and perhaps     
even enhanced, displaying a slightly bumpy feature  whereas the use     
of the CG scale produces a smoother metal-poor tail;    
    
\item     
a significant very metal-rich population with  $[Fe/H]>-0.4$ is clearly     
present, though the separation from the metal-rich side of the "central"     
component is sligtly less clear. We also note that a large fraction    
of these metal-rich objects populate the bin -0.4, -0.2, due the    
already mentioned different behaviour of the two metallicity    
scales in the very metal-rich regime.      
    
\item     
the percentage of the VMR stars varies from field-to-field as with     
the CG-scale;  G219 is confirmed to be peculiar.    
    
\end{itemize}    
    
In summary, the only possible (and very marginal) difference between     
the two sets of results is the slightly different appearance of the MDs     
in the metal-poor range, that could be interpreted as an indication     
for the existence of a possible secondary peak at $[Fe/H]_{ZW}$      
$\sim -1.5$ in a few fields. However, we do not attach much weight to     
this interpretation, since this might be due to, or enhanced by, a     
slight discontinuity  in the behavior of the RGB colors as a function     
of [Fe/H]$_{ZW}$  in the template grid, that occurs at [Fe/H]$_{ZW}$      
$\sim$ --1.4.  This discontinuity is not present in the CG scale.      
    
We note again that the choice of a different metallicity scale does not     
make any dramatic difference in the results, at least at the level of     
detail we believe is compatible with the intrinsic quality of the available      
data.

\subsection{ The distance}      
    
The panels (c) and (d) in Fig. 17 consider the effect of varying the M31     
distance modulus by $\pm 0.15$ mag on the MD of G64.      
This variation is rather large and must be considered an upper limit,     
yet the variation induced on the mean and mode of the metallicity     
distribution is only $\pm 0.07$ dex.     
    
Also the variation induced on the relative contributions of the three    
possible components defined by the three metallicity regimes (Sect. 6.2)    
is quite small.

\subsection{The total reddening}     
    
The effect of a total reddening variation of $\pm 0.05$ is not negligible,     
producing  a variation of $\pm$ 0.13 dex on the mean and mode of the     
MD,  and, consequently, would alter quite significantly the relative     
contribution of the smaller components (i.e. MPorY, VMR) to the global     
abundance distribution.     
However, the blue locus of the RGB, along with other    
estimates of the reddening toward M31, constrain the reddening of     
our fields well within this range.      
    
As already noted, since within the adopted approach the  MDs are    
fully drawn from the distributions of the intrinsic colors translated into    
MD via the adopted grid, it is unavoidable that any systematic shift     
of the colors correspondingly affects the MDs. And, since the     
relationship between color and metallicity is highly non-linear, even    
a small variation of the adopted reddening (at the level    
of a few hundredths of a magnitude) induces both a shift and a     
deformation of the MDs, especially in the metal-poor regime    
where the sensitivity to color variation is much larger.    
Note however that the median and mean metallicity estimates do not   
vary more than $\pm 0.2$ dex in response to a $\pm 0.05$ change   
in reddening.   
    
\subsection{Conclusions from the Tests}     
    
The results of the above tests can be summarized as follows:    
    
{\bf i)}     
With any plausible choice of distance modulus and total reddening,    
most of the stars (i.e. $\sim  60-80$\% of the total) lie in the   
range $\rm -1.4\le [Fe/H]\le -0.2$   and the {\em peak} of this main   
component of the  MD lies between $\rm [Fe/H]= -0.8$ and   
$\rm [Fe/H]=-0.4$, independently of the metallicity scale.     
    
{\bf ii)}     
Although the existence, as distinct populations, of two additional components     
at the  very metal-poor and  very metal-rich ends of the MDs may be     
debatable, the existence of very metal poor stars (also supported by the     
presence of a BHB population) and of very metal-rich stars seems     
to be quite firmly established.    
                         
{\bf iii)}     
As repeatedly noticed, the uncertainties in deriving abundances from     
photometry are still rather large, and settling the question    
of whether the abundance distributions are merely skewed or bimodal    
or even tri-modal will require much more accurate and reliable     
means of metallicity determinations, e.g. better photometric data     
coupled with a more reliable and extended grid of reference GC or the     
availability  of spectroscopic data for a wide sample of stars.

\section{Comparison with other MDs in the literature}    
    
Several authors have investigated M31 fields using both ground-based     
facilities and $HST$ observations.  We here compare our present results     
with those of the most recent previous analyses, noting     
that all of them work in the ZW metallicity scale.    
    
\begin{itemize}    
    
\item{\bf \citet{hof96}}    
    
These authors studied the halo fields near G302 and G312, located 32     
and 50 arcmin approximately along the SE minor axis, respectively.     
The reddening and distance modulus assumed for M31 were     
$E(B-V)=0.08 \pm 0.02$ and $(m-M)_0=24.3 \pm 0.1$.     
The RGBs were compared with the RGB ridge lines for three Galactic GCs      
(i.e. M15, NGC1851 and 47 Tuc) and two metal-rich fiducials selected from     
isochrones with age $t_0=13.8$ Gyr and [m/H]=--0.4 and 0.0.     
The resulting MDs show a spread in metallicity of $-2 \le [m/H] \le -0.2$     
with the majority of stars having $[m/H] \sim -0.6$.     
    
We have not analysed these fields, but we can compare these results with     
ours on halo fields such as G105 and G319.      
The $(V-I)_0$ color distribution of the RGBs at the approximate level     
I $\sim$ 22.45  (i.e. $M_I = -2.0$) seems to peak around 1.25 $\pm$ 0.2,     
at a cursory inspection of their Fig. 2 and 3;     
the resulting MDs are in very good agreement with our results,      
both in the shape and in the location of the MD peak at     
$[Fe/H] \sim -0.6$.     
    
\item{\bf \citet{svd}}    
     
The disk-dominated field near G272 was recently analysed by \citet{svd},     
who derived a MD for it. Their analysis is based on the assumptions     
that $E(B-V)=0.08$ and $(m-M)_0=24.5$ for M31.      
Their fiducial sequences are the RGB ridge lines of 6 Galactic globular     
clusters and one open cluster, of which only 47 Tuc is in common with our     
grid.      
    
The interpolation procedure, similar in principle to ours, was applied     
within a narrow range of magnitude along     
the RGB, i.e. I=22.65 $\pm$ 0.1 corresponding to  $M_I=-2.0 \pm 0.1$ mag.        
The $(V-I)_0$ color distribution within this strip     
(Fig. 6 in \citet{svd}) has a Gaussian shape peaking at $(V-I)_0 \sim 1.4$,     
whereas the color distribution we derive from our data under     
our assumptions peaks at $(V-I)_0 \sim 1.2$. This difference is largely     
accounted for by the different choice of reddening; differences in the absolute   
calibration (e.g., aperture corrections) may also contribute.   
    
The MD obtained by \citet{svd} is similar in shape to ours, with an extended    
metal-poor tail reaching [Fe/H] $\le$ --2, but their main peak at     
[Fe/H] $\sim$ --0.2 is definitely shifted by about 0.2 dex toward the     
metal-rich end. This is at least partly due to the quoted systematic     
difference in the color distribution.     
    
On this basis, these authors conclude that their Gaussian component peaking     
at $<[Fe/H]>=-0.22\pm0.26$ comprises $70$ \% of the total number of stars     
in the sample, and is attributed to the thick-disk population.     
    
We cannot compare this result homogeneously with ours because we     
have not fitted our MD with Gaussian components: however, we note that     
the fraction of very metal-rich stars with [Fe/H] $\ge$ --0.4 is $\sim$     
10\% in the ZW scale (see Fig. 18 and Table 5), which is not unusually     
large and is quite consistent with the position of G272 with respect to     
the major axis of M31. If indeed we are observing the thick-disk population     
in this field, its relative contribution is not overwhelming and is found     
also, and to a larger extent, in all the other fields located on the disk     
of M31 (see Sect. 9).

\item{\bf \citet{fj01}}    
    
The far outer disk field near G327 was recently investigated by    
\citet{fj01}. Under the standard assumptions of E(V--I)=0.10 and     
$(m-M)_0$=24.47, the RGB was compared with the fiducial ridge lines of three     
GGCs in the metallicity range --1.9 to --0.3. A best fit was performed at the     
luminosity level $M_I=-0.13\pm0.1 ~mag$, consistent with a predominantly     
old-to-intermediate age stellar population with [Fe/H] $\sim$ --0.7 plus     
a trace population of old metal-poor stars.    
     
In this field we also find a MD with a median value at [Fe/H] $\sim$ --0.66    
(ZW-scale), and only $\sim$ 20\% of stars with  [Fe/H] $<$ --0.8.

\item{\bf \citet{du01}}    
    
An outer halo field located 90 arcmin (i.e. 20 kpc) SE of the M31     
nucleus and roughly along the minor axis was studied by \citet{du01}      
using CFHT V,I data. A reddening value $E(V-I)=0.10\pm0.02$ was derived     
from the color distribution of the foreground Milky Way halo stars,     
and a distance modulus $(m-M)_0=24.47\pm0.12$ was derived from the     
luminosity of the RGB tip.     
    
A cursory inspection of their Fig. 7 indicates     
that the RGB stars at V $\sim$ 22.6 (corresponding to $M_I \sim -2$)     
populate a (V--I) color range between approximately 0.9 and 1.6 with     
a presumable accumulation between 1.2 and 1.4.     
    
The MD functions are derived by comparison with evolutionary tracks     
for 0.8 $M_{\odot}$ stars. The MD for the RGB magnitude interval      
20.6 $<$ I $<$ 22.5 is skewed toward the metal-poor side reaching      
almost [m/H]=--2.5, and could be well fitted by two Gaussian components     
peaking at [m/H]=--0.52 and --1.20 and including 60\% and 40\% of the     
total stellar population, respectively. The     
agreement with our analysis of the distant fields G105 and G319   
is quite satisfying (see Tables 4 \& 5).

\item{\bf \citet{rg02}}    
    
A spectroscopic study based on Keck-LRIS data of about 30 halo red giant     
stars in a field at R=19 kpc on the SE minor axis of M31 was     
done by \citet{rg02}.     
The MD they find for these halo giants spans more     
than 2 dex range with a mean/median value [Fe/H] $\sim$ --1.9 to --1.1     
(depending on calibration and sample selection). However, the high metallicity     
end of this distribution is poorly constrained by these data since the     
selection function for secure M31 members excludes $>$ 80\% of the     
giants in solar/super-solar metallicity range.     
    
No direct comparison is possible with our results, except for a general     
comment on the confirmation of the existence  of a well-detectable    
fraction of metal-poor stars in the outer halo of M31 (see Sect. 6.2).     
Our field G319 is within $\sim$ 30 arcmin of their field on the minor axis,     
and we do observe the MR population found in other fields.     
    
\end{itemize}

\section{Correlation of metallicity with position}    
    
\subsection{The wide-field survey by \citet{fer02}}    
    
In a very important  photometric survey of the halo and outer    
disk (appeared when the present analysis was nearly completed),     
\citet{fer02} have studied in great detail both the spatial density and    
metallicity variations (as inferred from color information,    
like in the present paper), covering an area of about 25 square     
degrees around M31. This is by far the most detailed and complete    
ground-based study carried out so far on the field population in M31.     
    
Since the CCD data were obtained with the    
2.5m INT-WFC at La Palma, the limiting magnitudes (with    
$S/N=5$) are V=24.5 and $i=23.5$, much brighter than those    
we have obtained from our $HST$ photometry. The shallower limits in     
magnitude are however amply balanced by the much wider     
area sampled. Therefore this survey and our work      
from $HST$ data are nicely complementary.     
    
Their conclusions, very convincingly supported by their fig. 2-7,    
are schematically:    
    
\begin{itemize}    
    
\item    
There is evidence for both spatial density and metallicity variations     
across the whole body of M31,    
which are often, but not always, correlated.     
    
\item    
Besides the known Ibata's stream, two other overdensities at large    
radii can be detected, close to the SW major axis, in the proximity    
of the very luminous GC G1 (called the "G1-clump"), and near the NE    
major axis, coinciding with and extending beyond the previously known    
"northern spur".    
    
\item    
The most prominent metallicity variations are found in two large    
structures in the southern half of the halo, the first one coinciding     
with the giant stellar stream found by \citet{iba01}, and the second one     
corresponding to a much lower stellar overdensity.     
Their metallicities are above the average value, which corresponds    
approximately  to that of 47 Tuc, i.e. $[Fe/H] \sim -0.7$.     
    
\end{itemize}    
    
Since the areas covered by the previous studies are very small    
compared to this wide survey, as correctly pointed out    
by Ferguson et al. there is no ground to identify significant     
conflicts between the results of previous studies on individual fields     
and those derived from this general panoramic study.     
This is especially true because of the clear detection    
of streams and substructures (both in star density and metallicity)    
allover M31, which has made somehow "unpredictable" the nature of    
the population one would expect to find on the basis of the mere      
X,Y location within M31.    
    
This fact has an obvious impact on the general interpretation     
we might derive from our analysis, as the lines of sight    
we have investigated, though quite numerous (16 + G1), could be    
not sufficiently representative of the global scenario.     
    
In this new light we have reconsidered the properties of our observed     
fields,  with particular attention to the  {\it map of the spatial and     
chemical substructures}  \citet{fer02} presented in their Fig. 7,     
and reported here in Fig. 20 where the locations of our fields     
are overplotted.       
 We anticipate that the results coming out from our deeper $HST$ fields    
suggest that indeed some of the considered lines of sight may be affected     
by the presence of streams and overdensities.

\subsection{Our 16+1 $HST$ fields}    
    
As reported in figures 14 and 15 and Table 4, a number of parameters     
have been derived from the metallicity distributions for each    
field. A straightforward application is to check first for any trend     
with position in the galaxy, recalling the already quoted caveat     
that only projected distances are available.     
    
\subsubsection{The (X,Y)--plane}    
     
Fig. $21a,b,c$ report the fractions of MPorY and VMR stars as a function of     
the X and Y coordinates and galactocentric distance R in arcmin,     
respectively. The error bars associated to each point are based on    
the square roots of the star numbers in each individual sample.    
    
Because of the large difference in the error bars     
associated to the different fields, as the respective population size       
ranges from about 10,000 stars in the inner fields down to less than 100    
in the outer halo fields, it is quite hard to assign a fair statistical     
significance to any trend one might see.    
Nevertheless, considering also the histograms presented in figures 14 and 15,     
some general comments can be made:     
    
\begin{itemize}    
   
\item    
The fraction of MPorY stars is substantially constant and independent    
of position. As already noted, this does not necessarily mean    
that there is a constant percentage of metal-poor halo stars in all the    
observed fields, because of the possible (and, at the present    
status, unknown) contamination by Young objects in the inner    
samples. However, metal-poor old stars are surely present in the    
halo, albeit in rather small quantity.      
   
 \item    
The inner fields (G87, G287) have a population with a higher     
average metallicity than all other fields, resulting from    
the combination of a higher than average VMR fraction and     
a (correspondingly) smaller MR fraction (remember that the     
total MPorY + MR + VMR = 1.0).    
This evidence is especially clear looking at the Y-plot (Fig. 21b).     
    
\item     
\citet{fer02} noted that the average metallicity    
of the SW fields (i.e. with negative Y and negative X) is slightly    
larger than the average metallicity estimated for the NE    
fields. Though the statistical significance is weak,     
this effect is perhaps visible also in Fig. 21a,b, as a   
slight enhancement of the mean fraction of  VMR    
stars in the $Y<0$ and $X<0$ regions.    
    
\item G219 and G319 have an unusually large fraction of metal-rich stars   
for their distance from the M31 center (90 and 70 arcmin respectively).   
This anomaly may be connected with their (projected) proximity to the   
\citet{fer02} stream.   
   
\end {itemize}    
    
In summary, the considerations reported above show that    
a simple description of the results making use of just    
the $HST$ fields and their projected location in the  X.Y-plane    
coupled with a "simple" description of the M31 outer disk and halo    
may be not fully capable to extract all the information    
potentially offered by the available data.    
    
This has prompted us to look at our result within     
the new scenario emerging from \citet{fer02} work.    
    
\subsubsection{HST fields and detected substructures}    
    
If we briefly rediscuss our data within the framework reported    
in Fig. 20 (where the fields are identified over the cartoon    
reproduced from Fig. 7 of Ferguson et al.), we can add some    
further notes:    
    
\begin{itemize}    
    
\item    
Several fields (e.g. G33, G219, G105, G58, G64, G108)     
are located or projected on the "giant stream" first identified by     
\citet{iba01} and now confirmed by \citet{fer02}.     
    
Some of these fields (see Table 4,5) show a very high fraction of VMR stars,    
with a clear peak in the MD.  The G219 field presents one of the  
best examples of this case.  G219 is located well off the plane  
but lies precisely on the     
detected stream; the field has an anomalously large  
fraction ($\sim 20$\%) of very metal rich stars.     
    
Though one cannot conclude for sure that the existence of these     
VMR stars in most of these fields has to be ascribed to the    
possible physical connection with the "giant stream", it    
seems conceivable that the two facts may be related.    
On the other hand, there are also fields that are      
located/projected on the stream which do not show a     
similarly evident "excess" of VMR stars (see f.i. G58, G64, G105).    
    
\item    
The MD of G1 may also be worth of a short comment within the    
new scenario. In fact, even though it is a distant field (projected  
some 32 kpc from the nucleus) and with a total population of measured stars of    
only 44 objects, its MD shows (see Fig. 16) a weak indication    
for a second metal-rich peak (at $[Fe/H] \sim -0.2$, with    
F(VMR) $\sim 10$\%.     
    
Since, as shown in the map in Fig. 20, the region called by    
Ferguson et al. the "G1 clump" denotes a possible substructure    
(at least in star density) within the area they surveyed, it    
may be of some relevance to investigate further the existence    
and origin of VMR objects in this field.    
    
\end {itemize}    
    
In summary, the re-analysis of the properties of the MDs    
within the framework provided by \citet{fer02}     
does not yield a {\it fully} satisfactory  ordering and   
detailed explanation for all the fields investigated in the     
present study. However, some (otherwise unusual) results are     
nicely fitting (or, at least, are compatible with) a scenario   
that accounts for the existence of significant substructures     
within M31. These may add an important piece of information for   
a detailed description  of the formation and evolution of this galaxy,   
including the possible interaction with close companions such as M32   
\citep{bek02,cho02}.    
    
\subsection{Large-scale trends}    
    
In the previous sections we have carried out    
and discussed the analysis of the MDs we have obtained     
from the histograms shown in figures 14 and 15 assuming     
that, independently of the physical reasons or adopted models for M31,    
one could identify two features or discontinuities,    
at $[Fe/H]=-1.4$ and $[Fe/H]=-0.2$, to separate  the stars    
into three main groups.    
We are quite confident that such an approach is meaningful and      
the considerations listed above add some support to this.     
    
An alternative and more schematic approach could be simply to     
consider two groups of stars obtained by     
dividing the total MDs for instance  at $[Fe/H]=-0.8$.     
    
This value for the cut has been chosen  to identify roughly   
what we call hereafter the ``Metal-Rich Population'' (MRP)     
from the ``Metal-Poor Population'' (MPP)  on the basis    
of the already quoted (weak) evidence  of a small discontinuity   
in the MDs at that bin-border    
(deduced from inspecting the histograms, see f.i. G33, G322,     
G108, G58, G351, G219, G327; see also Fig.~12).     
     
Table 5 reports for each field the values so obtained from the histograms     
for the fraction of the Metal-Poor Population , F(MPP), and the     
Metal-Rich Population, F(MRP), the Y-coordinate, and the average (median)    
metallicity of the total sample.     
    
Fig. 22 a,b shows the plots of F(MPP) and of the median     
metallicity vs. the absolute Y-coordinate, respectively.     
    
As said, the contribution of stars from the spheroid is expected to grow     
faster along the Y-direction because of the inclination of the M31     
disk with respect to the plane of the sky.    
     
In fact, F(MPP) grows very rapidly from 0.10 at $|Y_{arcmin}| \sim 0$, to     
$\sim 0.45$ at $|Y_{arcmin}| \sim 17$, showing a good correlation     
between the involved variables. However, from $|Y_{arcmin}| \sim 20$     
to $|Y_{arcmin}| \sim 70$, F(MPP) remains nearly constant at    
F(MPP)$\sim 0.4$, with a larger  scatter.     
Note that the two ``outliers''  with low values    
of F(MPP) at quite large Y are G219 and G351, both located    
on or near the "giant stream" (see Fig. 20).    
A similar behaviour is shown by the median metallicity (Fig. 22b)    
    
The trend displayed in figures 22a and 22b is quite clear:     
the MRP dominates the inner region but the relative importance of     
the MPP population rapidly grows going far from the visible disk.     
However, in the explored range of $|Y_{arcmin}|$,     
{\em the Metal-Poor population never becomes dominant, Metal-Rich stars remain     
the major component of the stellar mix}.    
    
In summary, these plots  suggest     
that, besides the existence of important substructures, there is also     
evidence of  a more regular pattern in the inner parts of the galaxy,    
possibly testifying     
of a relatively homogeneous and ordered formation process,     
later perturbed by merging events.

\section{Summary and Conclusions}    
   
We have analysed HST-WFPC2 images of 16 fields in M31 at a wide range of   
distances from the galactic center and plane.   
These fields have been observed in the F555W (V) and F814W (I) filters  
and reach a roughly uniform depth.  
   
The color-magnitude diagrams of these fields are generally dominated by   
a red giant branch and a populous red clump. The RGB has a strong descending tip,   
indicative of a metal rich population. In the outermost fields a   
blue Horizontal Branch is also detected.   
When present, the blue HB represents about 15\%\ of the total HB   
population, roughly estimated from  the fraction of   
metal poor giants.   
   
In fields superposed on the disk,   
a blue plume of main sequence stars is also identified.       
We report a simple analysis of the blue plume population,   
which shows that the star formation history of the disk has been spatially   
inhomogenous during the last 0.5 - 1 Gyr.   
   
We have obtained photometric metallicity distributions from RGB stars   
by interpolation on a grid of empirical globular cluster RGB templates.   
   
The most robust result of     
the present $HST$ survey is the evidence that the metal-rich population,    
with  $[Fe/H] \sim -0.6$, is the major component of the    
stellar mix everywhere in the $10 \le R_{arcmin} \le 130$ range.    
A  minor metal-poor component (with $[Fe/H]<-1.4$) is also ubiquitous.    
The old stellar population is remarkably uniform across the disk,  
outer halo fields, and in the proposed tidal stream.  This uniformity  
is the dominant feature of the old stellar population.   
    
This basic result is not new    
\cite[see, e.g.,][and references therein]{du01},    
but coupled with data shown by    
the very wide mapping carried out by \citet{fer02},    
fixes on solid grounds the conclusion that the stellar population   
of the M31 spheroid differs substantially from that of the    
Milky Way and is almost an order of magnitude more metal rich, on average.    
In this respect, we confirm also the finding of \citet{rich96}, that    
the metal rich population is present even at 150 arcmin     
($\sim$ 35 kpc) from the nucleus, at the location of the globular     
cluster G1.    
   
An obvious explanation for the origin of the metal rich    
halo of M31 is that the population formed in connection with the    
outflow of metal-enriched winds, perhaps associated with the formation    
of the spheroid.  Metal rich populations in halos are clearly    
widespread \cite[e.g. NGC 5128; ][]{sor97,hh99,hh02}.    
There is evidence for the outflow of metal enriched winds in    
Lyman break galaxies, and for near-solar abundances in these    
galaxies, at high redshift \citep{ste96}.    
The stars formed from this material might well live to comprise    
the population II halo.   The outflow of enriched material will not   
affect the shape of the abundance distribution (the classic Simple   
Model form is retained) but the yield (mean abundance of the whole   
galaxy) does decline \citep{hart76}.   
   
The increasing fraction of the "Metal-Poor Population"    
(i.e. with $[Fe/H]< -0.8$) with distance from the plane Y     
suggests a dissipational origin for the regions that are closer   
to the galactic plane.   This evidence is compatible with the   
hypothesis that a "homogeneous"  formation event may have   
occurred in the very early stages of the  M31 history.    
    
On the other hand,    
the image of the M31 halo illustrated by \citet{iba01}     
and \citet{fer02} shows (for the metal-rich stars taken    
as a {\em global} population) a quite flattened, disk-shaped distribution,    
associated with the outskirts of the inner disk. It is possible thus    
that the metal rich population might be associated with a proto disk or   
flattened halo.  The large number of stars involved would strongly favor  
the notion that we study the  
{\it old } disk, and not just some trace population such    
as the thick disk.  So we have this alternative scenario, where   
the dominant metal-rich  population might be associated with   
a meta-disk extending perhaps more the 20 kpc, while   
the metal-poor population might be  associated  with the spheroid.     
The two populations -- and the corresponding structures -- would have     
different evolutionary histories. The old stellar disk would be      
much larger than previously believed, extending out to a few degrees     
from the nucleus of M31 and dominating the stellar mix also   
along lines of sight very distant from the center of the galaxy   
\citep{fj01,fer02}.    
It is also possible that the inner disk and bulge had common, or closely   
tied, formation histories.    
  
In addition to this, the sub-structures (``streams'' and ``clumps'')   
clearly detected by \citet{iba01,fer02} and supported by our results   
might be related to the phenomena of merging or interaction with   
satellites. In this case the metal rich component might be connected with   
the tidal disintegration of a companion (\citet{fer02}),   
a likely candidate being M32, perhaps involving a good deal of its mass   
\citep{bek02,cho02}.    
  
New data are now required to choose among the scenarios presented   
here.  Deeper HST imaging, some of which is in progress, may constrain  
the actual age distribution of stars in some regions of the halo.  
Large scale spectroscopic surveys of halo stars have just begun, and  
have the potential to determine what fraction of the M31 halo might  
have originated from the tidal disruption of satellites.  Our first  
round of large-scale HST imaging gives evidence that the old ``halo'' stellar  
population is more metal rich than that of the Milky Way, with surprisingly  
little variation in the properties of the old stellar population,  
even for fields ranging up to galactocentric distances exceeding 30 kpc.  
The challenge now will be assemble a data set for the Galaxy and M31   
powerful enough to constrain their origins.

\acknowledgements{    
We are indebted to Helmut Meusinger for kindly providing us with the plates     
of the Tautenburg Schmidt Telescope, and to Roberto Merighi for help    
in drawing some figures. Useful discussions with  Gisella   
Clementini, Carlo Corsi, George Djorgovski, Francesco Ferraro, Wendy    
Freedman, Puragra Guhathakurta, Monica Tosi    
are also kindly acknowledged.   Support for Michael Rich's activities   
on proposal GO-6671 was provided by NASA trhough a grant from the Space   
Telescope Science Institute, which is operated by the Association of   
Universities for Research in Astronomy, Inc., under NASA contract   
NAS 5-2655. Grants: ASI J/R/35/00, MURST MM02241491-004.}

%

%
  
\clearpage  
\begin{table}
\begin{center}
\caption{Location of the observed fields.} 
\begin{tabular}{lrrrrrr}
\hline  \hline
\multicolumn{7}{c}{} \\
  G & Bo & X$_{arcmin}$ & Y$_{arcmin}$ & R$_{arcmin}$ & E(B-V) & F$_{sph}$ \\  
\multicolumn{7}{c}{} \\
\hline  \multicolumn{7}{c}{} \\
 G1 &   &$-149.54 $&$ 29.88 $&$152.50 $&$0.11 $&$ >0.9$\\
 G11&293&$ -61.72 $&$ 43.91 $&$ 75.75 $&$0.11 $&$ >0.9$\\
 G33&311&$ -57.58 $&$  1.24 $&$ 57.59 $&$0.11 $&$ 0.07$ \\
 G58&~~6&$  -6.80 $&$ 27.37 $&$ 28.20 $&$0.11 $&$ 0.70$ \\
 G64&~12&$ -10.65 $&$ 23.01 $&$ 25.35 $&$0.11 $&$ 0.65$ \\
 G76&338&$ -44.15 $&$ -8.84 $&$ 45.03 $&$0.14 $&$ <0.1$ \\
 G87&~27&$ -26.41 $&$  1.00 $&$ 26.43 $&$0.14 $&$ 0.18$ \\
G105&343&$ -57.57 $&$-29.81 $&$ 64.83 $&$0.11 $&$ 0.80$ \\
G108&~45&$   7.38 $&$ 20.21 $&$ 21.52 $&$0.11 $&$ 0.60$ \\
G119&~58&$ -28.88 $&$-10.06 $&$ 30.59 $&$0.14 $&$ <0.1$ \\
G219&358&$ -64.79 $&$-58.32 $&$ 87.17 $&$0.11 $&$ >0.9$ \\
G272&218&$  17.13 $&$-16.88 $&$ 24.05 $&$0.14 $&$ <0.1$ \\
G287&233&$  35.44 $&$ -0.26 $&$ 35.45 $&$0.14 $&$ 0.12$ \\
G319&384&$ -21.18 $&$-68.89 $&$ 72.07 $&$0.11 $&$ 0.99$ \\
G322&386&$  61.64 $&$ -4.49 $&$ 61.80 $&$0.11 $&$ 0.07$ \\
G327&   &$ 130.01 $&$  5.10 $&$130.11 $&$0.11 $&$ >0.9$ \\
G351&405&$  63.14 $&$-53.71 $&$ 82.90 $&$0.11 $&$ >0.9$ \\
\hline
\end{tabular}
\end{center}
\end{table}
  
\begin{table*}[h]
\begin{center}
\caption{Observations. }  
\begin{tabular}{lccllcl}  
\hline  \hline
\multicolumn{7}{c}{} \\
Name & R.A. (2000)   & Dec. (2000)   &  
  Date &Proposal  & P.I. & \multicolumn{1}{c}{Filters (t$_{exp} [s]$) } \\  
\multicolumn{7}{c}{} \\
\hline
\multicolumn{7}{c}{} \\  
 G11  & 0 36 20.78 & 40 53 36.60 & Feb  3,2000   &GO 6671  & Rich	& F814W (5400), F555W (5300)\\  
 G33  & 0 39 33.75 & 40 31 14.36 & Feb 26,1999   &GO 6671  & Rich	& F814W (5400), F555W (5300)\\  
 G64  & 0 40 32.79 & 41 21 44.65 & Aug 17,1999   &GO 6671  & Rich	& F814W (5400), F555W (5300)\\  
 G76  & 0 40 58.83 & 40 35 47.32 & Jan 11,1999   &GO 6671  & Rich	& F814W (5400), F555W (5300)\\  
 G87  & 0 41 14.61 & 40 55 51.12 & Aug 16,1999   &GO 6671  & Rich	& F814W (5400), F555W (5300)\\  
G119  & 0 41 53.01 & 40 47 08.63 & Jun 13,1999   &GO 6671  & Rich	& F814W (5400), F555W (5300)\\  
G287a & 0 44 41.97 & 41 43 56.40 & Sep 26,1999   &GO 6671  & Rich       & F814W (5400), F555W (5300)\\  
G287b & 0 44 41.97 & 41 43 56.40 & Sep 26,1999   &GO 6671  & Rich       & F814W (5400), F555W (5300)\\  
G319  & 0 46 21.92 & 40 17 00.01 & Feb 28,1999   &GO 6671  & Rich	& F814W (5400), F555W (5300)\\  
G322  & 0 46 26.94 & 42 01 52.94 & Jan 10,1999   &GO 6671  & Rich	& F814W (5400), F555W (5300)\\  
G327  & 0 49 38.91 & 43 01 13.85 & Jun 19,1999   &GO 6671  & Rich	& F814W (5400), F555W (5300)\\  
      &  	   &		 &		 &	   &	        &			    \\  
G272  & 0 44 51.40 & 41 19 16.00 & Jan 22,1995   &GO 5420  & Fusi Pecci & F814W(10800), F555W (3800)\\  
G351  & 0 49 58.10 & 41 32 17.00 & Jan 18/19,1995&GO 5420  & Fusi Pecci & F814W(10800), F555W (3800)\\  
      &            &             &               &         &            &                           \\  
 G58  & 0 40 26.79 & 41 27 27.72 & Feb 15,1994   &GTO 5112 & Westphal	& F814W (2000), F555W (2000)\\  
G108  & 0 41 43.26 & 41 34 20.76 & Feb 15,1994   &GTO 5112 & Westphal	& F814W (2000), F555W (2000)\\  
G105  & 0 41 43.17 & 40 12 22.76 & Feb 15,1994   &GTO 5112 & Westphal	& F814W (2000), F555W (2000)\\  
G219  & 0 43 17.81 & 39 49 13.53 & Feb 15,1994   &GTO 5112 & Westphal	& F814W (2000), F555W (2000)\\  
\hline
\end{tabular}
\end{center}
\end{table*}
\clearpage
\begin{table}[p]
\begin{center}
\caption{Grid of Galactic Globular Clusters template for metallicity  
determinations in the \citet{zw84} and \citet{cg97} scales}  
\begin{tabular}{lllcc}  
\hline  \hline
\multicolumn{5}{c}{} \\
   &[Fe/H]$_{ZW}$&[Fe/H]$_{CG}$&E(B-V)&$(m-M)_{0}$\\               
\multicolumn{5}{c}{} \\
\hline  \multicolumn{5}{c}{} \\
NGC6341~~M92 &$ -2.24$&$  -2.16$&$0.02$&$14.74$\\
NGC6205~~M13 &$ -1.65$&$  -1.39$&$0.02$&$14.38$\\
NGC5904~~M5  &$ -1.40$&$  -1.11$&$0.03$&$14.31$\\
NGC104~~47Tuc&$ -0.70$&$  -0.70$&$0.04$&$13.29$\\
NGC6553      &$ -0.34$&$  -0.16$&$0.84$&$13.44$\\
NGC6528      &$ -0.23$&$~~~0.07$&$0.62$&$14.35$\\
\hline
\end{tabular}
\end{center}
\end{table}
  
\begin{table*}[h]
\begin{center}
\caption{Metallicity Distributions in the \citet{cg97} scale:   
population (in percentage)  of individual metallicity bins  
(step 0.2 dex) for each observed field, and for all (coadded)    
 fields. }  
{\scriptsize 
\begin{tabular}{lcrrrrrrrrrrrrrrcc}  
\hline  \hline
\multicolumn{18}{c}{} \\
Name&$<-2.5$&$-2.5$&$-2.3$&$-2.1$&$-1.9$&$-1.7$&$-1.5$&
$-1.3$&$-1.1$&$-0.9$&$-0.7$&$-0.5$&$-0.3$&$-0.1$&$ 0.1$&$.2-.27$&$>.27$ \\  
\multicolumn{18}{c}{} \\
\hline  \multicolumn{18}{c}{} \\
G1  &0.0&2.3&0.0&0.0&0.0&0.0&2.3&11.4&11.4&22.7&20.5& 4.5& 9.1& 9.1& 2.3&2.3&2.7\\  
G11 &0.0&1.1&0.0&0.0&0.6&0.6&1.1& 2.2& 9.0&17.4&24.7&23.0& 9.0& 9.0& 1.1&0.0&1.2\\  
G33 &1.3&0.2&0.2&0.4&0.6&0.7&0.4& 3.5& 5.8& 8.2&15.2&22.5&16.1&19.0& 4.8&0.5&0.6\\  
G58 &0.4&0.1&0.5&0.5&0.9&1.1&1.2& 5.9&11.2&15.5&25.6&23.7& 9.5& 3.3& 0.5&0.0&0.0\\  
G64 &0.6&0.2&0.2&0.4&0.4&0.8&0.5& 6.0& 8.6&14.7&21.5&25.6&11.5& 7.3& 1.3&0.1&0.5\\  
G76 &3.6&0.3&0.6&0.7&0.9&0.9&0.9& 3.7& 5.3& 8.5&13.6&21.8&15.7&16.7& 4.4&0.6&1.7\\  
G87 &0.4&0.0&0.2&0.2&0.3&0.3&0.4& 1.7& 3.2& 5.2&10.3&16.5&15.1&26.2&13.2&2.3&4.6\\  
G105&0.2&0.0&0.0&0.7&0.4&0.2&1.1& 5.0& 8.8&15.3&21.9&25.8&12.3& 6.8& 0.9&0.2&0.4\\  
G108&1.2&0.2&0.4&0.4&0.5&0.9&0.8& 5.0& 9.0&14.0&22.2&24.4&12.0& 7.7& 1.0&0.1&0.3\\  
G119&3.9&0.3&0.8&0.9&1.2&1.0&1.3& 5.2& 7.8&10.4&14.8&22.1&14.6&11.6& 2.8&0.5&0.8\\  
G219&0.8&0.0&0.0&0.0&0.8&1.6&2.4& 2.4& 2.4& 7.1&17.3&26.0&18.9&18.1& 2.4&0.0&0.0\\  
G272&4.1&0.3&0.6&1.1&1.5&2.0&1.6& 7.9&10.5&13.2&17.3&20.7&11.0& 6.4& 1.3&0.2&0.4\\  
G287&0.5&0.1&0.2&0.3&0.3&0.4&0.4& 2.3& 4.4& 7.9&13.9&24.3&18.6&20.7& 4.3&0.6&0.8\\  
G319&0.0&0.0&2.4&0.0&0.0&1.2&2.4& 4.8&11.9&13.1&17.9&11.9&13.1&11.9& 6.0&0.0&3.6\\  
G322&1.8&0.2&0.3&0.5&0.7&0.3&0.6& 3.4& 6.2&10.0&17.2&21.9&16.5&15.7& 3.7&0.3&0.5\\  
G327&0.0&0.0&2.4&0.0&0.0&0.0&0.0& 1.2& 8.3&10.7&33.3&19.0&13.1& 9.5& 1.2&0.0&1.2\\  
G351&0.8&0.0&0.8&0.0&0.8&0.0&0.0& 5.9& 7.6& 9.3&21.2&26.3&16.1& 5.1& 2.5&0.8&2.5\\  
ALL &1.7&0.2&0.4&0.5&0.7&0.7&0.7& 3.7& 6.0& 9.2&15.0&21.4&15.0&16.8& 5.5&0.9&1.7\\  
\hline
\end{tabular}   
}
\end{center}
\end{table*}
\clearpage
\begin{table*}[h]
\begin{center}
\caption{Parameters derived from the Metallicity Distributions in the  
\citet{cg97} scale. Columns:   
(1) name of the field,  
(2) fraction of stars with $\rm [Fe/H]<-1.4$ [Metal-Poor or Young -- MP or Y],  
(3) with $\rm -1.4<[Fe/H]<-0.2$ [Metal-Rich --MR],  
(4) with $\rm [Fe/H]>-0.2$ [Very Metal-Rich -- VMR]  
(5) fraction of stars with $\rm [Fe/H]<-0.8$ [Population Metal-Poor --PMP],  
(6) fraction of stars with $\rm [Fe/H]>-0.8$ [Population Metal-Rich --PMR],  
(7) average metallicity ($\rm [Fe/H]_{ave}$)   
(8) associated standard deviation,   
(9) median metallicity,   
(10)associated semi-interquartile interval.  
The corresponding figures in the \citet{zw84} scale are   
partially reported in Fig. 18 and 19.    }  
\begin{tabular}{lccccccccc}  
\hline  \hline
\multicolumn{10}{c}{} \\
Name&$Fr$(MP or Y)&$Fr$(MR)&  
$Fr$(VMR)&$Fr$(PMP)&$Fr$(PMR)&[Fe/H]$_{av}$&$\sigma_{av}$&
[Fe/H]$_{med}$&$\sigma_{med}$ \\   
\multicolumn{10}{c}{} \\
\hline  \multicolumn{10}{c}{} \\
G1   &0.045&0.795&0.136&0.500&0.477&-0.79&0.49&-0.82&0.26\\  
G11  &0.034&0.854&0.101&0.320&0.669&-0.68&0.38&-0.64&0.20\\  
G33  &0.026&0.713&0.242&0.200&0.781&-0.52&0.42&-0.48&0.25\\  
G58  &0.043&0.915&0.038&0.369&0.627&-0.75&0.37&-0.67&0.22\\  
G64  &0.025&0.878&0.087&0.317&0.672&-0.68&0.37&-0.63&0.22\\  
G76  &0.043&0.686&0.218&0.218&0.729&-0.55&0.46&-0.49&0.26\\  
G87  &0.013&0.520&0.418&0.114&0.837&-0.35&0.39&-0.28&0.25\\  
G105 &0.024&0.891&0.079&0.315&0.678&-0.67&0.35&-0.63&0.22\\  
G108 &0.031&0.866&0.087&0.311&0.673&-0.68&0.38&-0.63&0.22\\  
G119 &0.056&0.749&0.149&0.289&0.664&-0.65&0.47&-0.57&0.28\\  
G219 &0.047&0.740&0.205&0.165&0.827&-0.53&0.38&-0.50&0.22\\  
G287 &0.017&0.713&0.256&0.163&0.568&-0.48&0.37&-0.44&0.24\\  
G272 &0.071&0.805&0.079&0.386&0.824&-0.76&0.46&-0.67&0.28\\  
G319 &0.060&0.726&0.179&0.357&0.607&-0.67&0.49&-0.63&0.30\\  
G322 &0.026&0.713&0.242&0.224&0.753&-0.56&0.41&-0.52&0.25\\  
G327 &0.024&0.857&0.107&0.226&0.762&-0.65&0.39&-0.63&0.17\\  
G351 &0.017&0.864&0.085&0.246&0.720&-0.63&0.38&-0.56&0.22\\  
ALL  &0.032&0.702&0.233&0.220&0.746&-0.54&0.43&-0.50&0.27\\  
\hline
\end{tabular}
\end{center}
\end{table*}

  
  

\begin{figure*}
\centering
\includegraphics[width=17cm]{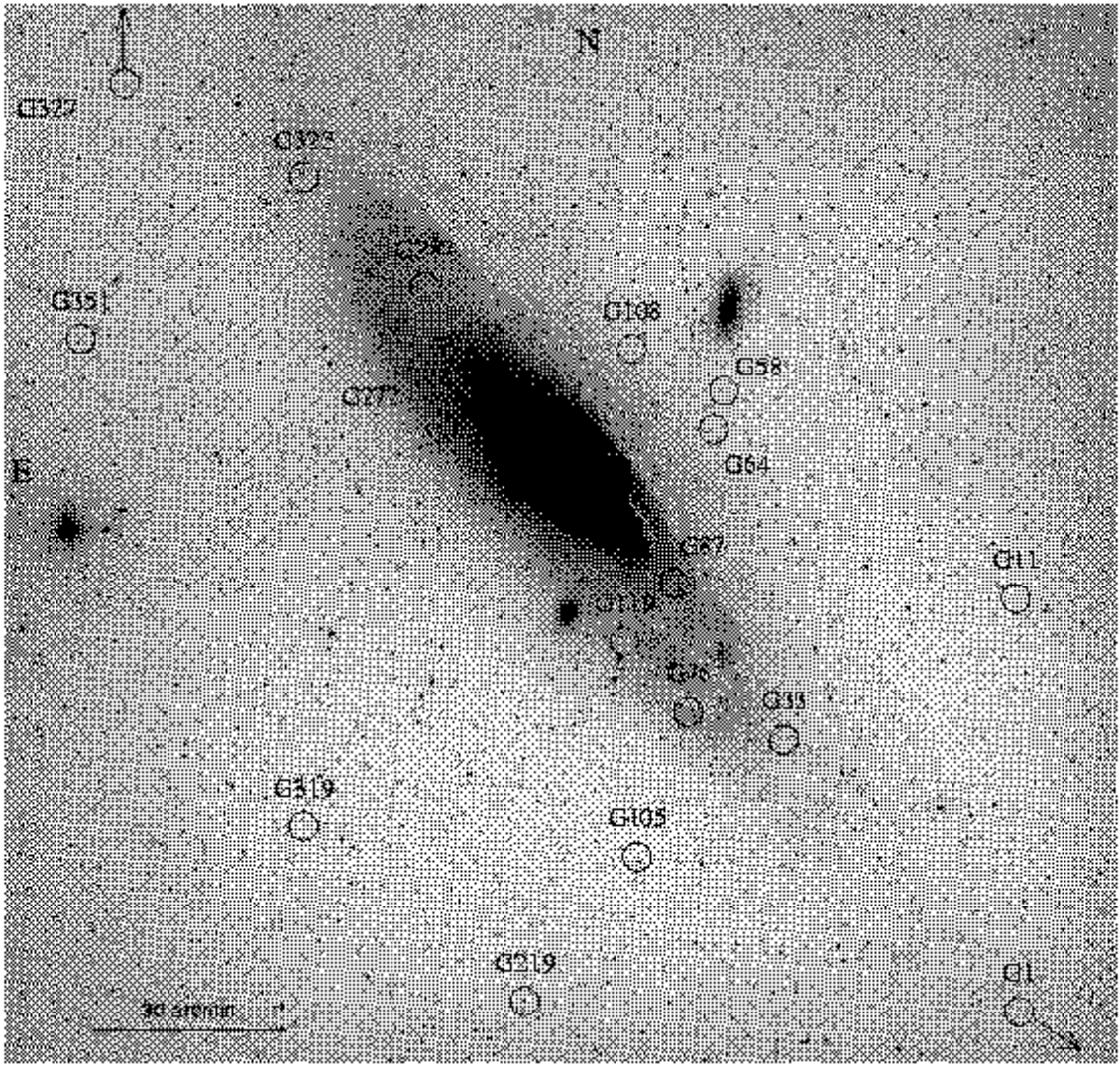}  
  \caption{The positions of the observed fields overplotted on an image of M31.}  
\end{figure*}

\clearpage  
  
\begin{figure*}  
\includegraphics{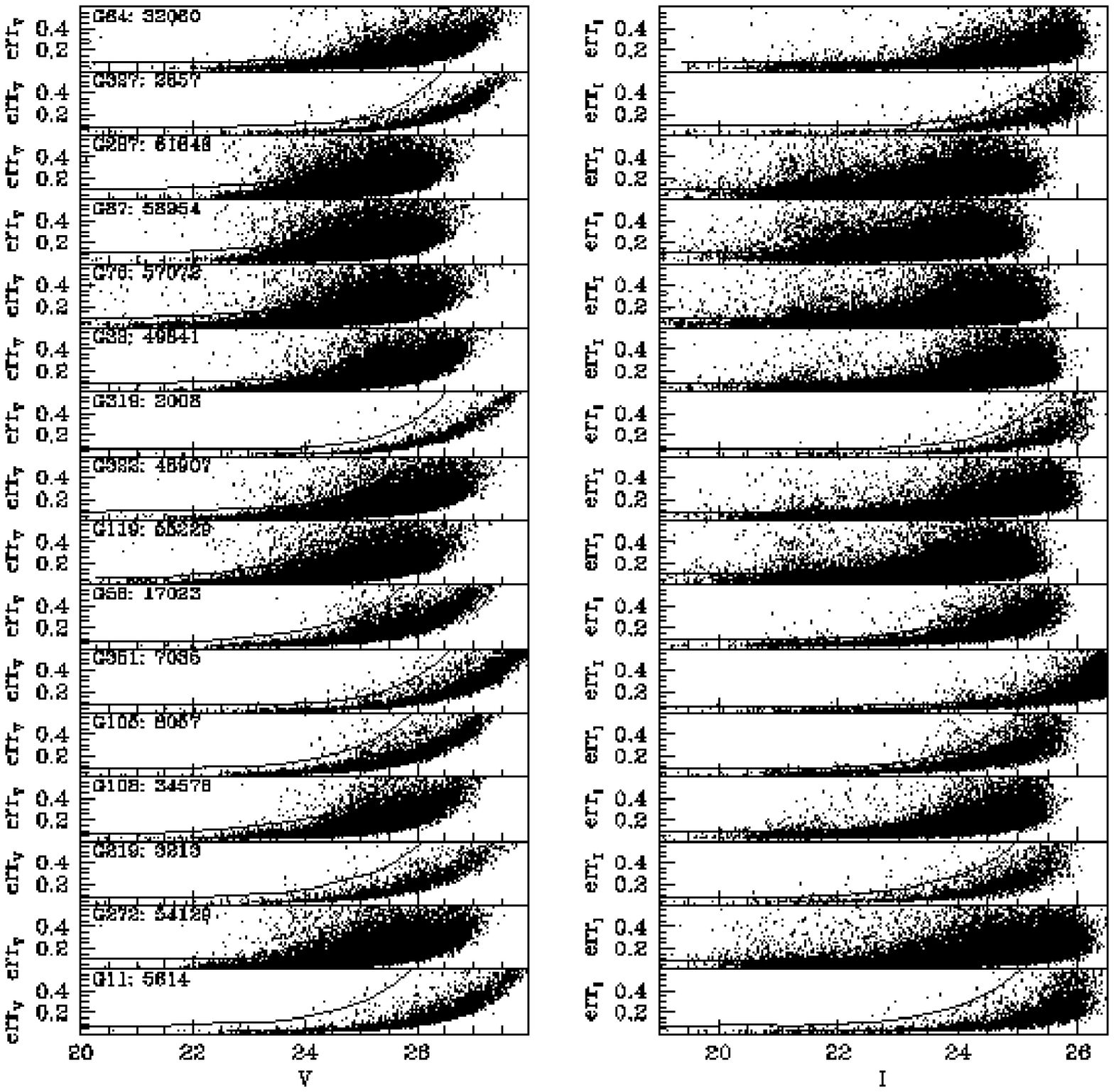}  
\caption{Photometric errors as a function of V and I magnitude.  
The lines represent the treshold for three times the average error.  
The points above the lines are excluded from the final sample.}   
\end{figure*}  
  
\clearpage  
  
\begin{figure*}  
\includegraphics[width=17cm]{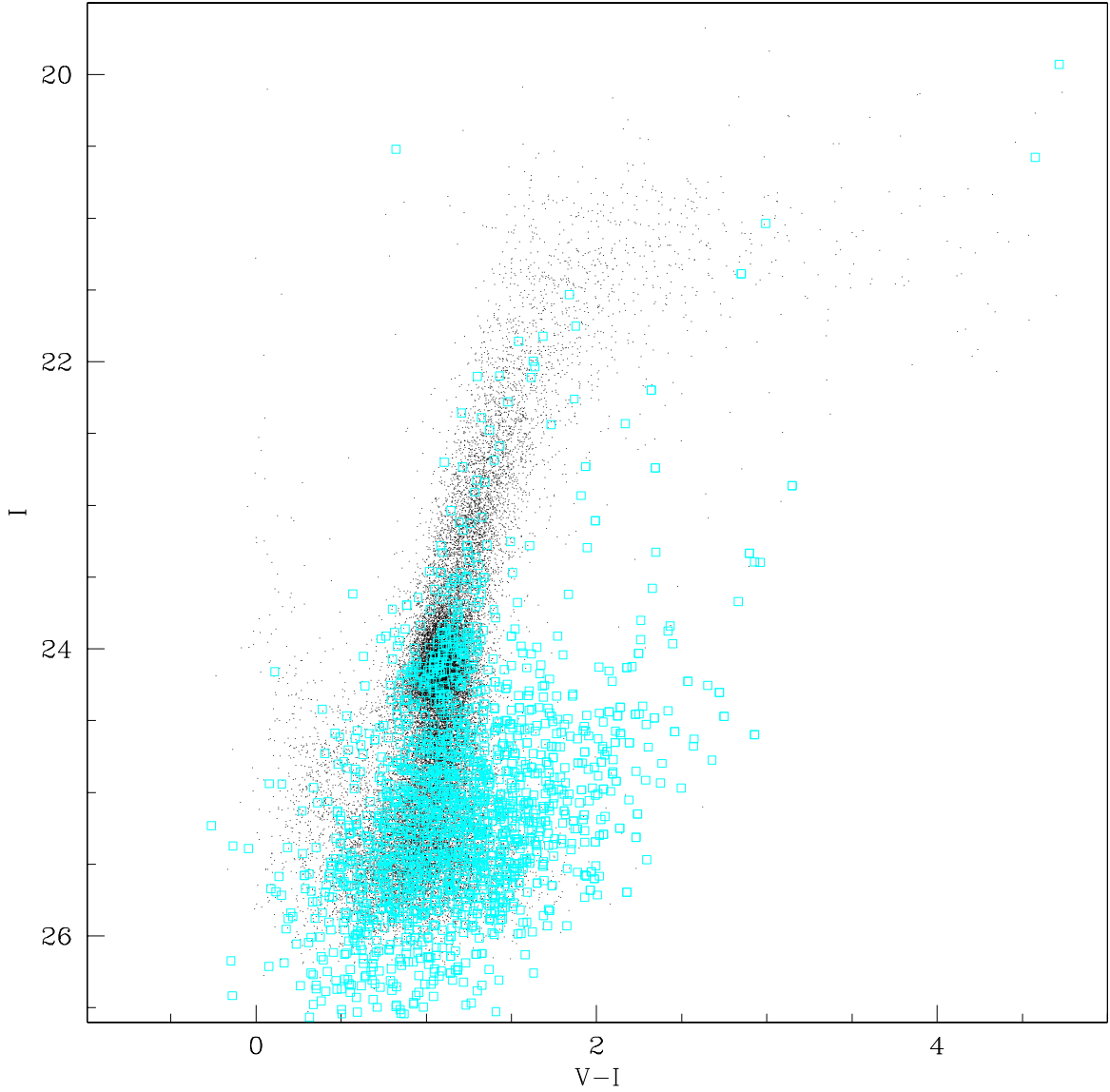}  
\caption{The manually removed contaminating sources (open squares)  
from the entire sample superposed on the Color Magnitude Diagram of the field  
G64. Most of the sources redder than $(V-I)\sim 1.5$ are galaxies or spurious  
stars from the decomposition of extended galaxies, most of the  
sources bluer than this figure are spurious stars from spikes and/or coronae  
of heavily saturated stars}   
\end{figure*}  
  
\clearpage  
  
\begin{figure*}  
\includegraphics[width=17cm]{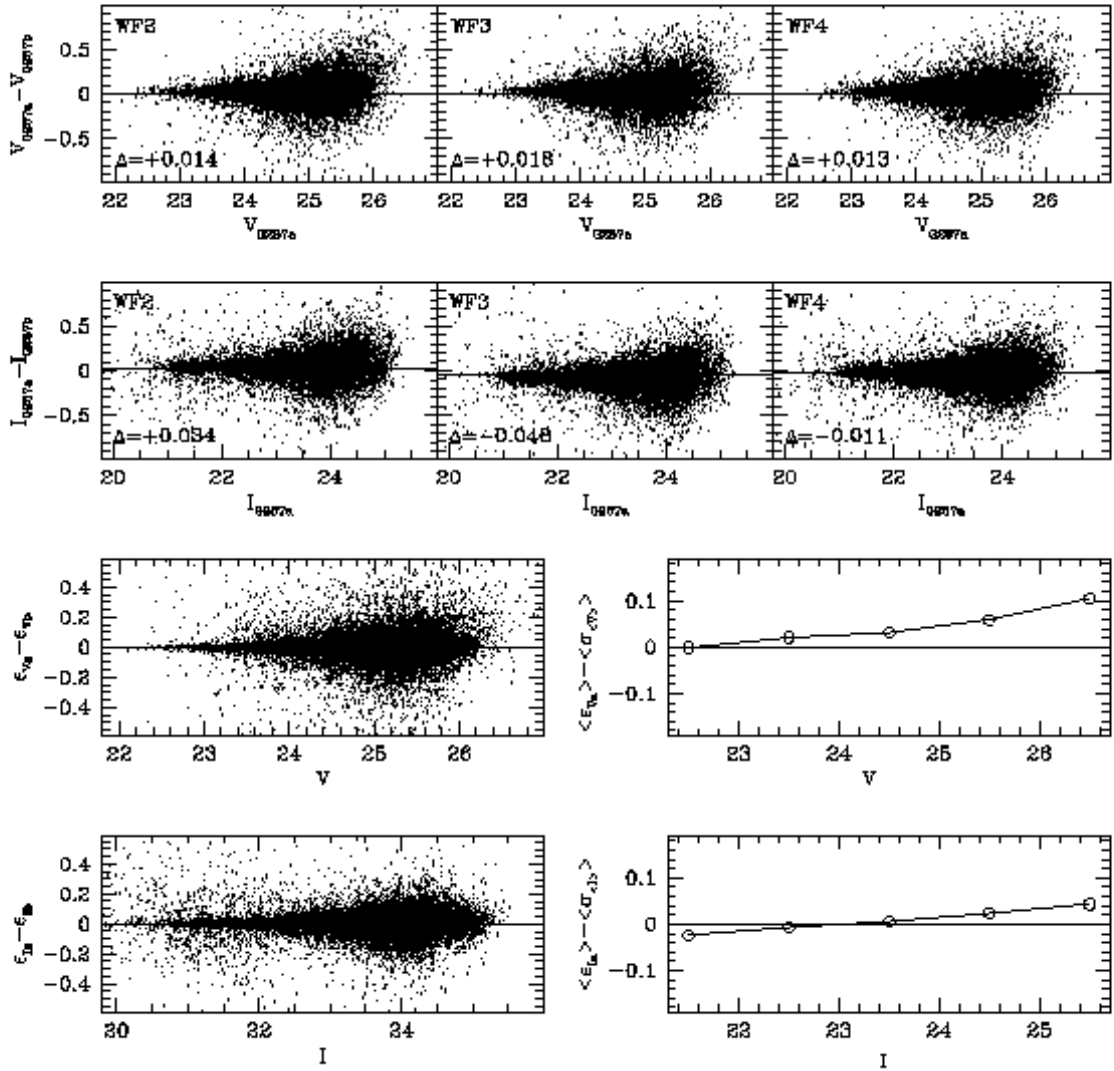}  
\caption{Comparison between the photometry of the stars in common  
in the fields G287a and G287b. In the upper two panels the difference in the   
final calibrated magnitudes is shown for the three WF cameras independently   
(first panel: V magnitudes, second panel: I magnitudes). In the lower two  
panels are reported the differences between the errors as computed by DoPHOT  
and the error on the mean obtained by the two repeated measures of the stars in  
common, versus magnitude in the corresponding filter (third panel: V, fourth  
panel I. The differences has been averaged over 0.5 mag boxes. The errors  
provided by DoPHOT are good estimates of the true photometric uncertainties.}   
\end{figure*}  
  
\clearpage  
  
\begin{figure*}  
\includegraphics[width=17cm]{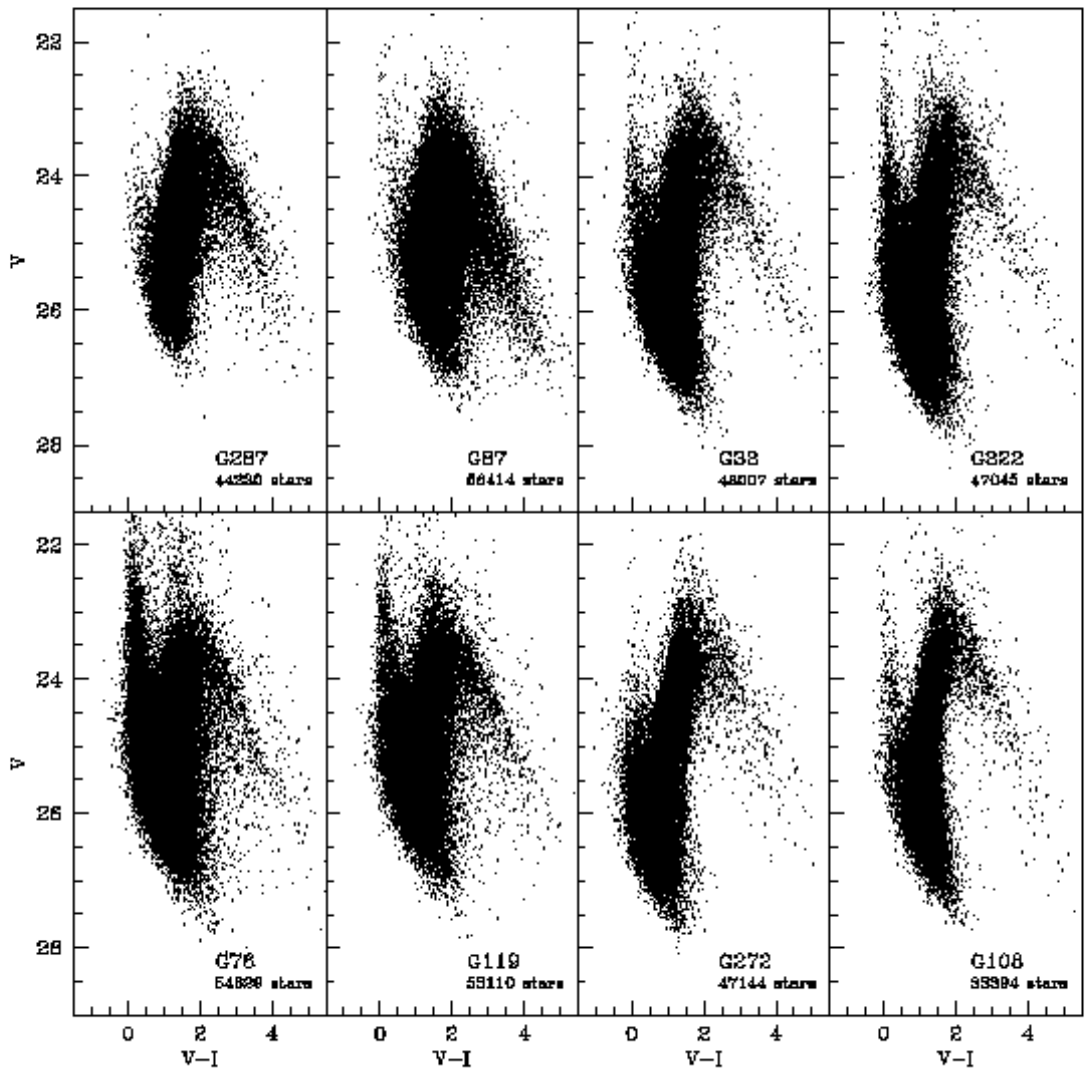}  
\caption{The (V,V-I) Color Magnitude Diagrams for the  
fields: G287, G87, G33, G322, G76, G119, G272, G108. In this figure and in  
figure 6 the diagrams are shown in order of increasing distance from the major  
axis, with the only exception of G327 (see text).}   
\end{figure*}  
  
\clearpage   
  
\begin{figure*}  
\includegraphics[width=17cm]{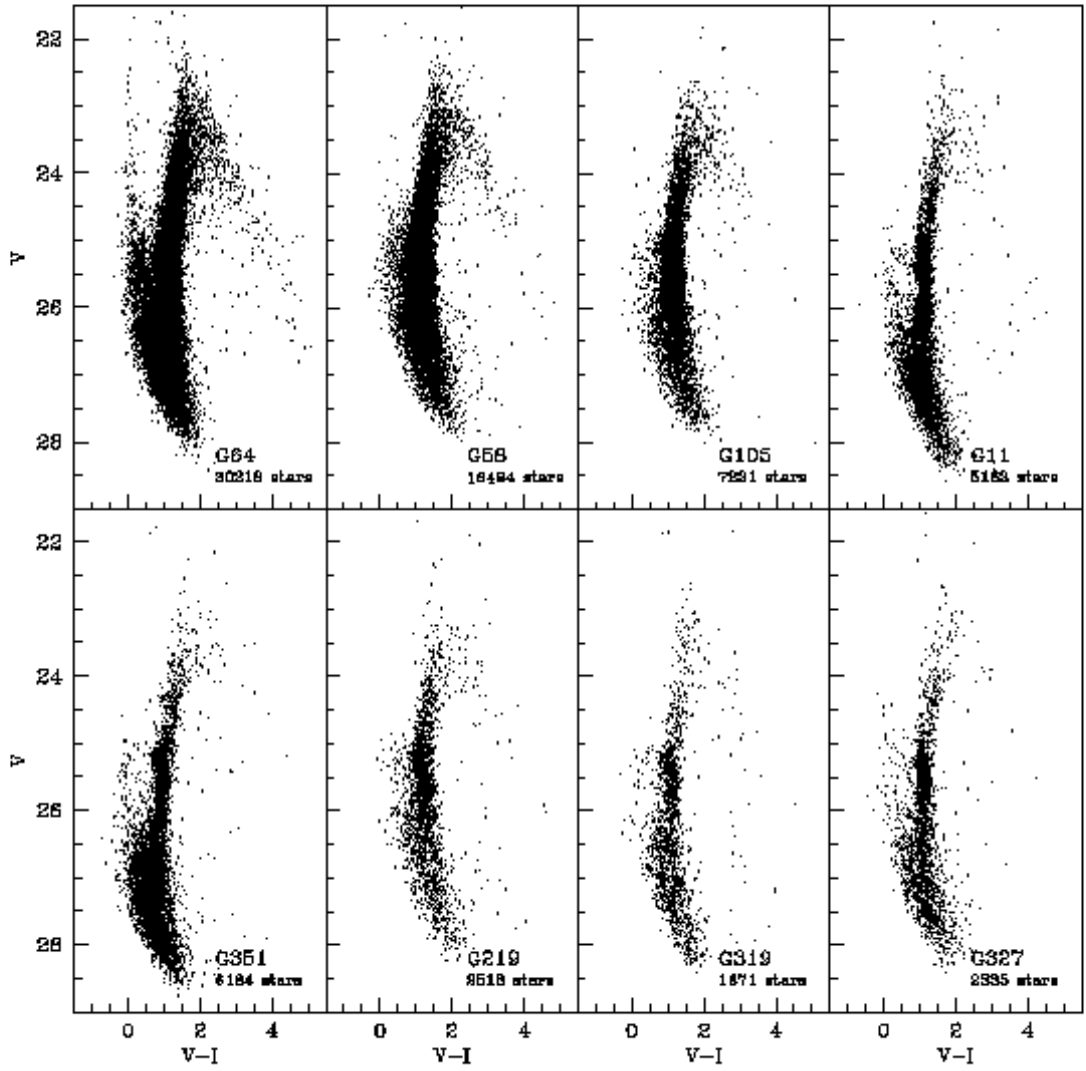}  
\caption{The (V,V-I) Color Magnitude Diagrams for the  
fields: G64, G58, G105, G11, G351, G219, G319, G327. See the caption of   
Fig. 5.}   
\end{figure*}  
  
\clearpage

\begin{figure*}  
\includegraphics[width=17cm]{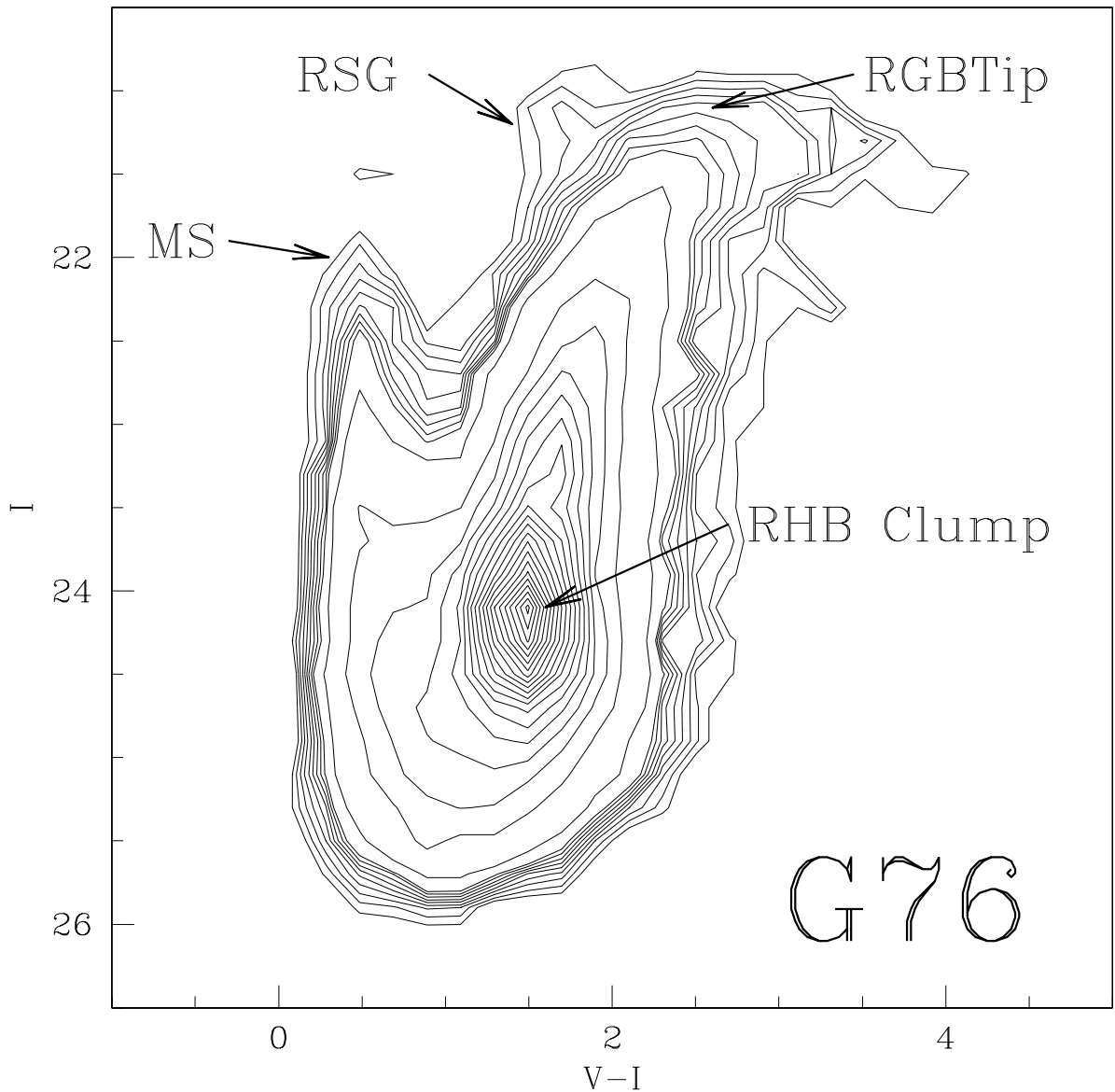}  
\caption{The CMD of the G76 field represented as an isodensity  
contour plot (Hess diagram) for greater clarity.  The main features  
described in the text are indicated.   
The outermost countour correspond to a density  
of 10 stars per $0.1 ~mag \times 0.1 ~mag$ box. The step between subsequent   
contours is 5 up to the 8th contour. The 9th contour corresponds to  
75 stars per $0.1 ~mag \times 0.1 ~mag$ box and after this contour the step is  
100. The innermost contour is at 2200 stars per   
$0.1 ~mag \times 0.1 ~mag$, while the box sampling the peak of the RHB Clump   
has a density of 2334 stars.}   
\end{figure*}  
  
\clearpage   
  
\begin{figure*}  
\includegraphics[width=17cm]{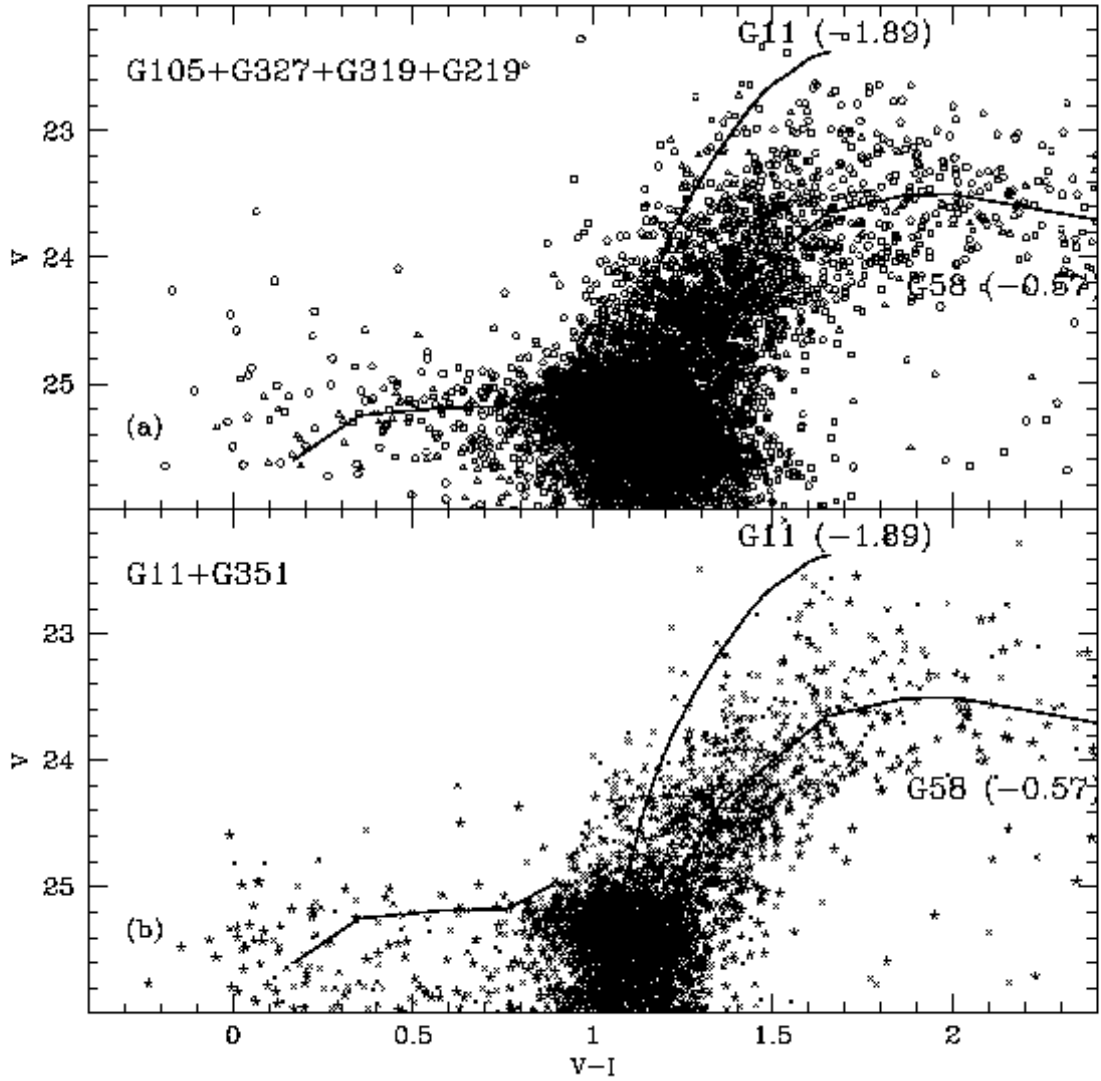}  
\caption{CMD for the coadded halo fields zoomed to show the details 
of the HB morphology (see text).   
The lines superimposed to the   
plot are the HB fiducial for M68, shifted by $\Delta$V=9.44 and   
$\Delta$V-I=0.09 to match the blue HB of the  
M31 fields, and the RGB ridge lines of the M31 globular clusters G11   
([Fe/H] = -1.89) and G58 ([Fe/H] = -0.57).  
Panel (a) the CMDs of G105, G327, G319 and G219. Stars from different
fields are marked with different symbols (G105: pentagons; G327: circles; G319:
triangles; G219: squares). 
Panel (b) the CMDs of G11 and G351 (G11: $\times$s;
G351: stars). The BHB of the CMDs presented in panel (b) is contaminated by an
apparent blue plume that is not present in the CMDs of panel (a).   
}   
\end{figure*}  
  
\clearpage  
\begin{figure*}  
\includegraphics[width=17cm]{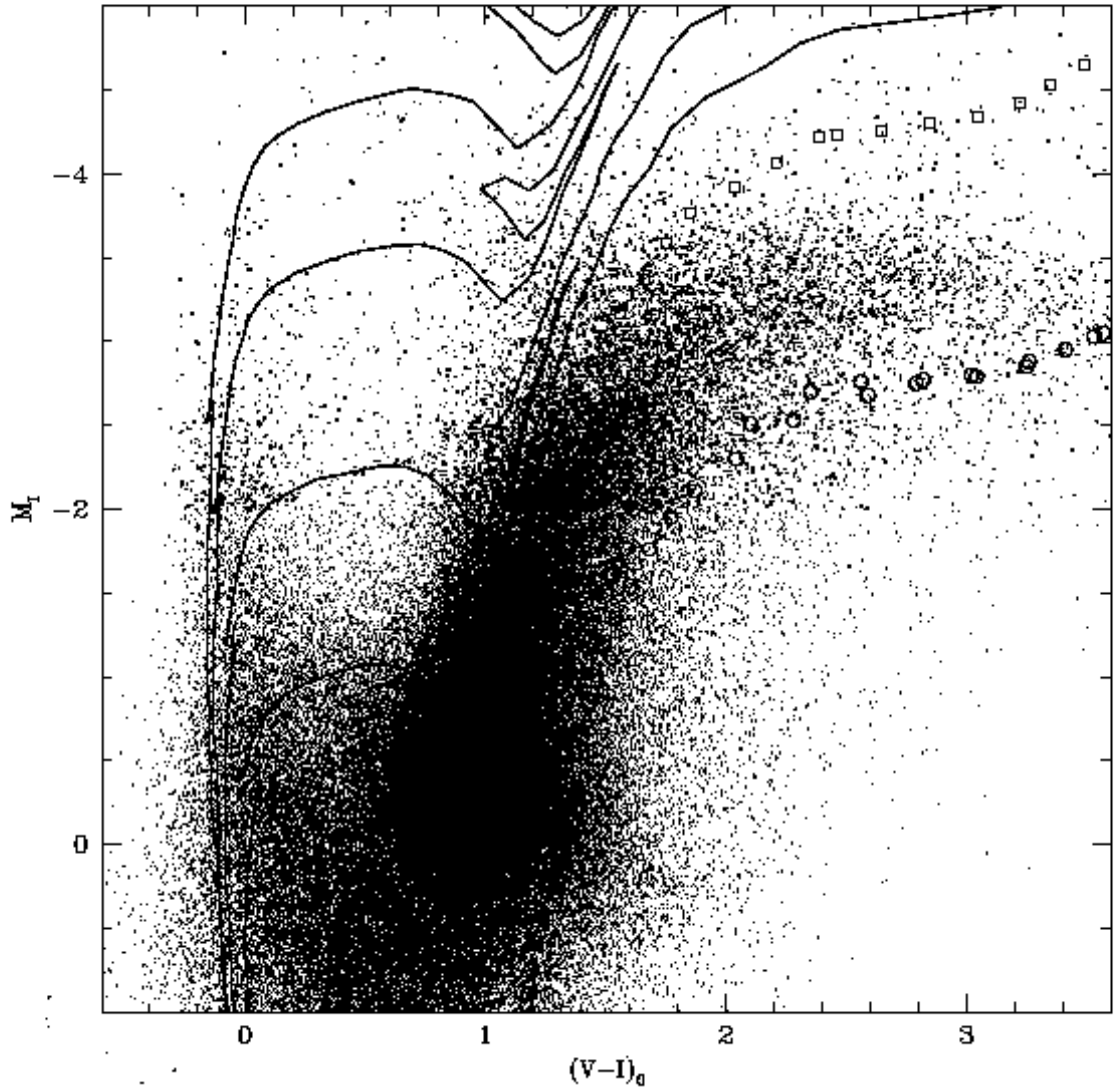}  
\caption{CMD of all the 154784 stars from the disk-dominated fields  
G76, G119 and G322, reported to the absolute plane $[M_I,(V-I)_0]$. Isochrones  
at $[Fe/H]=0.0$ and $Y=0.28$, from the set of \citet{bert94} are superimposed on  
the diagram. The continuous lines are isochrones of age t = 60, 100, 200 and  
400 Myr, from top to bottom. Open squares and circles correspond to isochrones  
of t = 1 and 12 Gyr, respectively. The stars brighter and fainter than   
the $M_I=-2$  threshold are shown as  
points of different thickness to allow easier recognition of both the  
densely populated features in the lower part of the CMD (e.g. the HB  
Clump) and the sparse bright features (e.g. the upper MS and the red plume  
of RSG stars}   
\end{figure*}  
\clearpage   
  
\begin{figure*}  
\includegraphics[width=17cm]{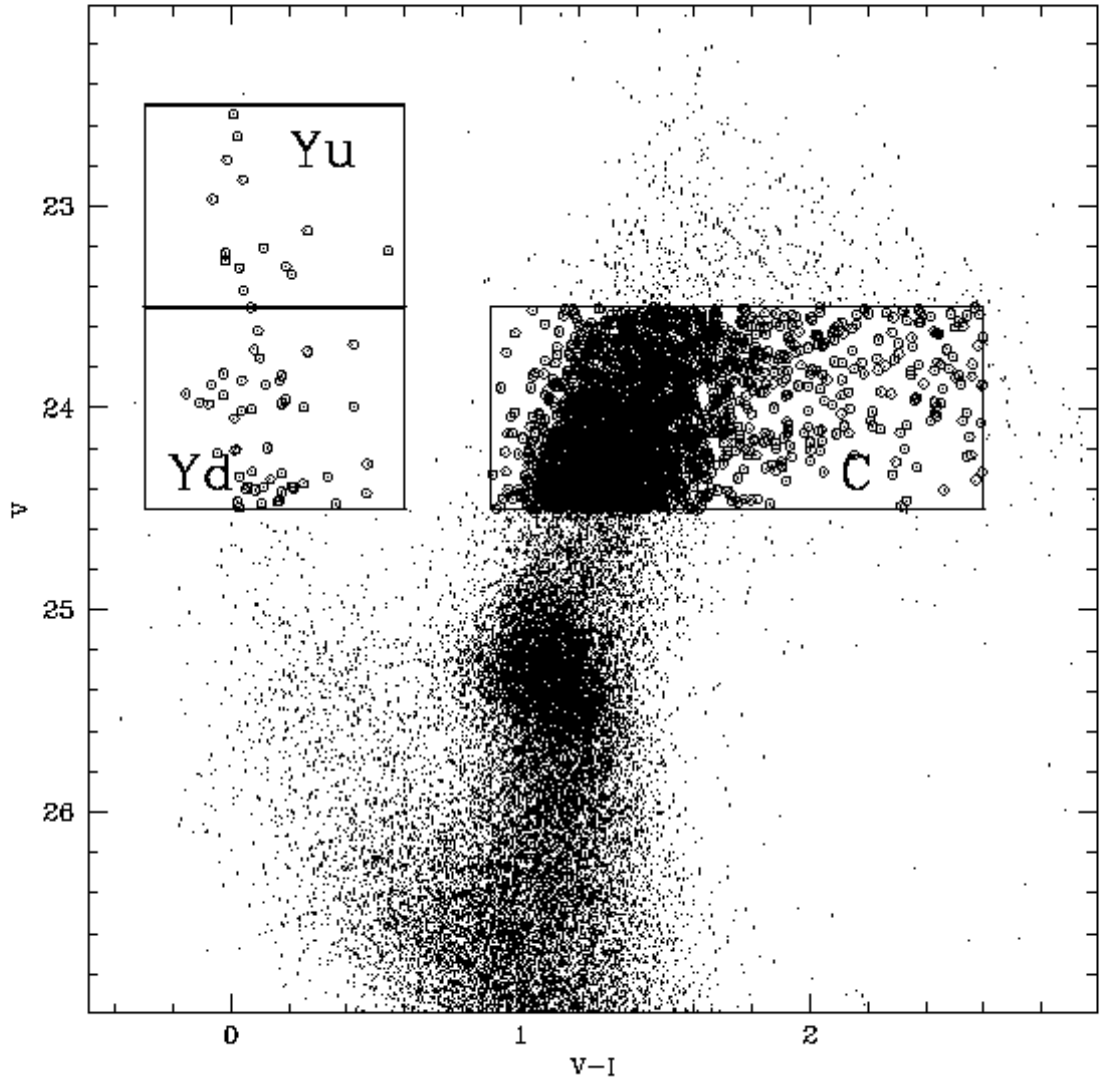}  
\caption{The population boxes defined to study the YMS are  
evidenced in the [V,(V-I)] CMD of the G64 field. The boxes marked with {\bf Yu}  
and {\bf Yd} samples different part of the Main Sequence, while the box marked  
with {\bf C} has been defined for normalization purposes. The stars falling into  
the indicated boxes has been indicated with open circles.}   
\end{figure*}  
  
\clearpage  
  
\begin{figure*}  
\includegraphics[width=17cm]{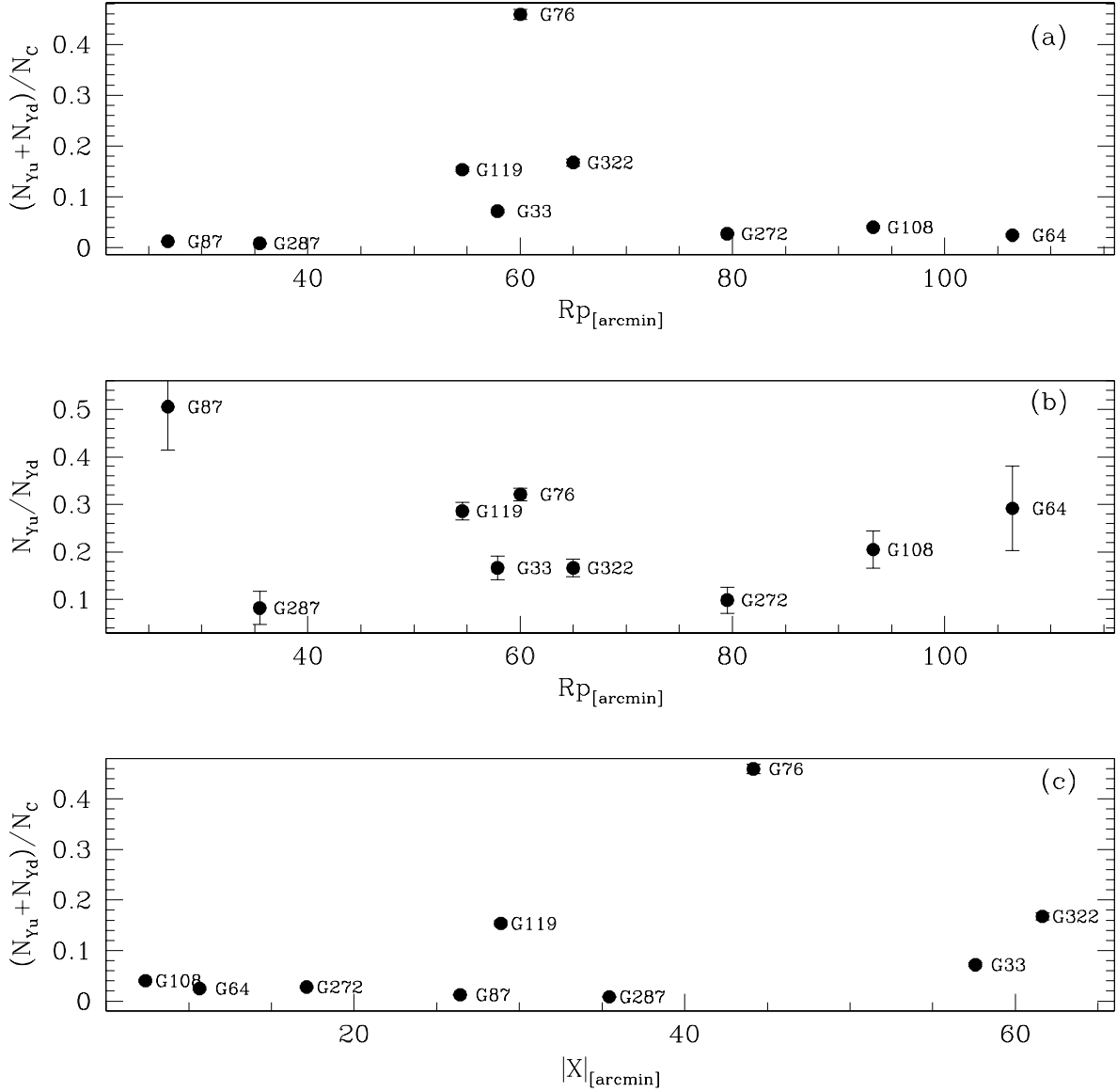}  
\caption{YMS (young main sequence) indices versus deprojected distance   
from the center  
of the galaxy (panels a and b), and versus absolute distance along the major   
axis (panel c). Where no error bars are seen, they are smaller than 
the symbol size. 
Only the fields with a substantial YMS population   
(e.g. disk fields) are plotted.  The upper and lower panels plot the 
total fraction of stars younger than $\sim 0.5$ Gyr as a function of 
distance from the nucleus, while the center panel shows the fraction 
of young stars as a function of galactocentric distance.  Note that 
while the G76 field has the strongest young population, it is the 
sparsely populated G87 field that has the youngest main sequence.  The 
G76 field coincides with a ring of general enhanced star formation 
10 kpc from the nucleus.}   
\end{figure*}  
  
\clearpage  
  
\begin{figure*}  
\includegraphics[width=17cm]{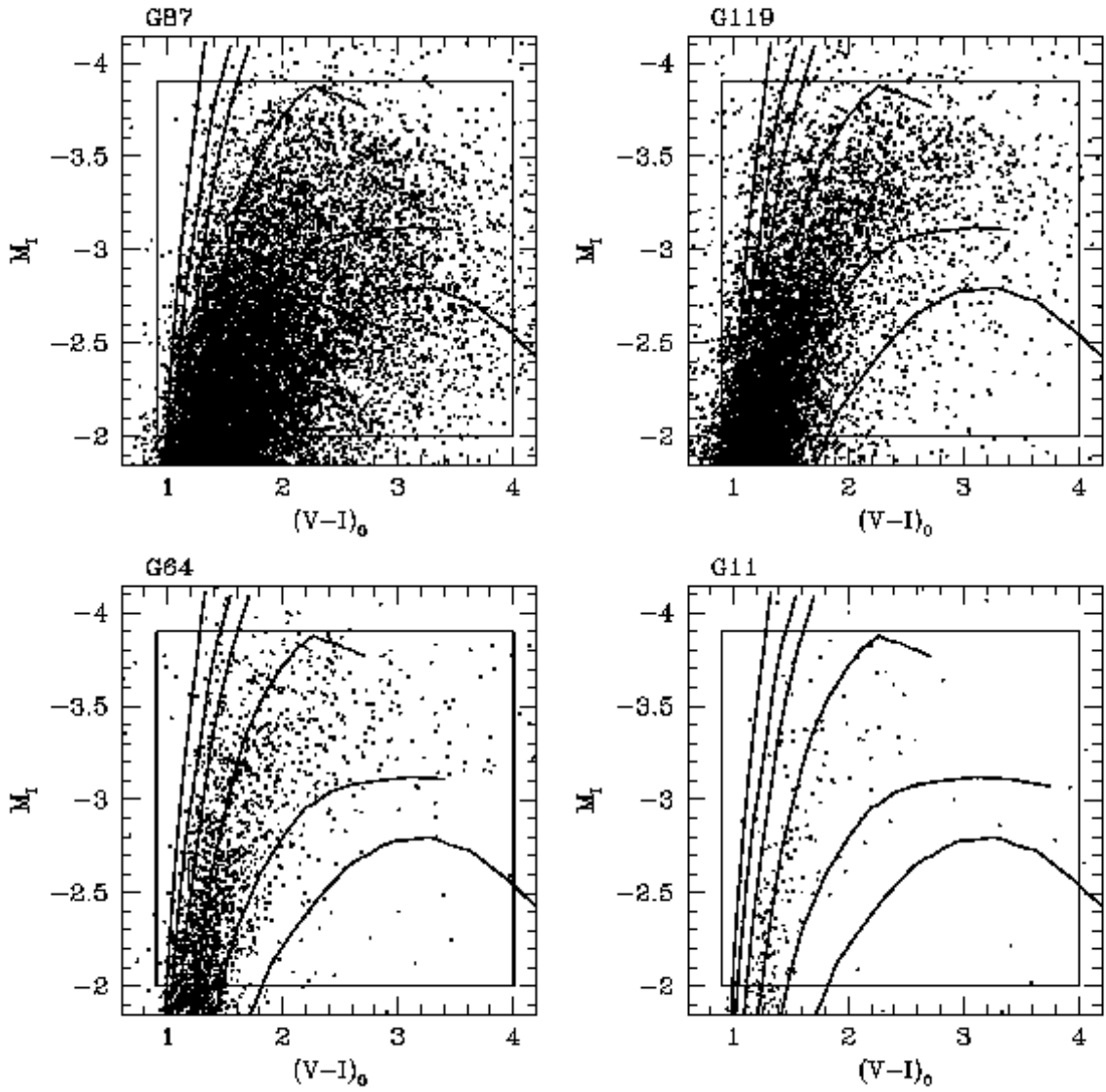}  
\caption{CMDs of the upper RGB in the absolute plane,  
for four fields, illustrating the full range of crowding in the  
data set.  
The ridge lines of template galactic globular clusters are superimposed to each   
plot. From left to right (CG metallicity scale):  
NGC 6341 ($\rm [Fe/H]=-2.16$), NGC 6205 ($\rm [Fe/H]=-1.39$), NGC 5904   
($\rm [Fe/H]=-1.11$) and NGC 104 ($\rm [Fe/H]=-0.70$), NGC 6553  
($\rm [Fe/H]=-0.16$)   
and NGC 6528 ($\rm [Fe/H]=-0.07$). The inner frame encloses the stars whose   
metallicities are determined using the   
interpolating scheme described in Fig. 13. }   
\end{figure*}  
  
\clearpage  
  
\begin{figure*}  
\includegraphics[width=17cm]{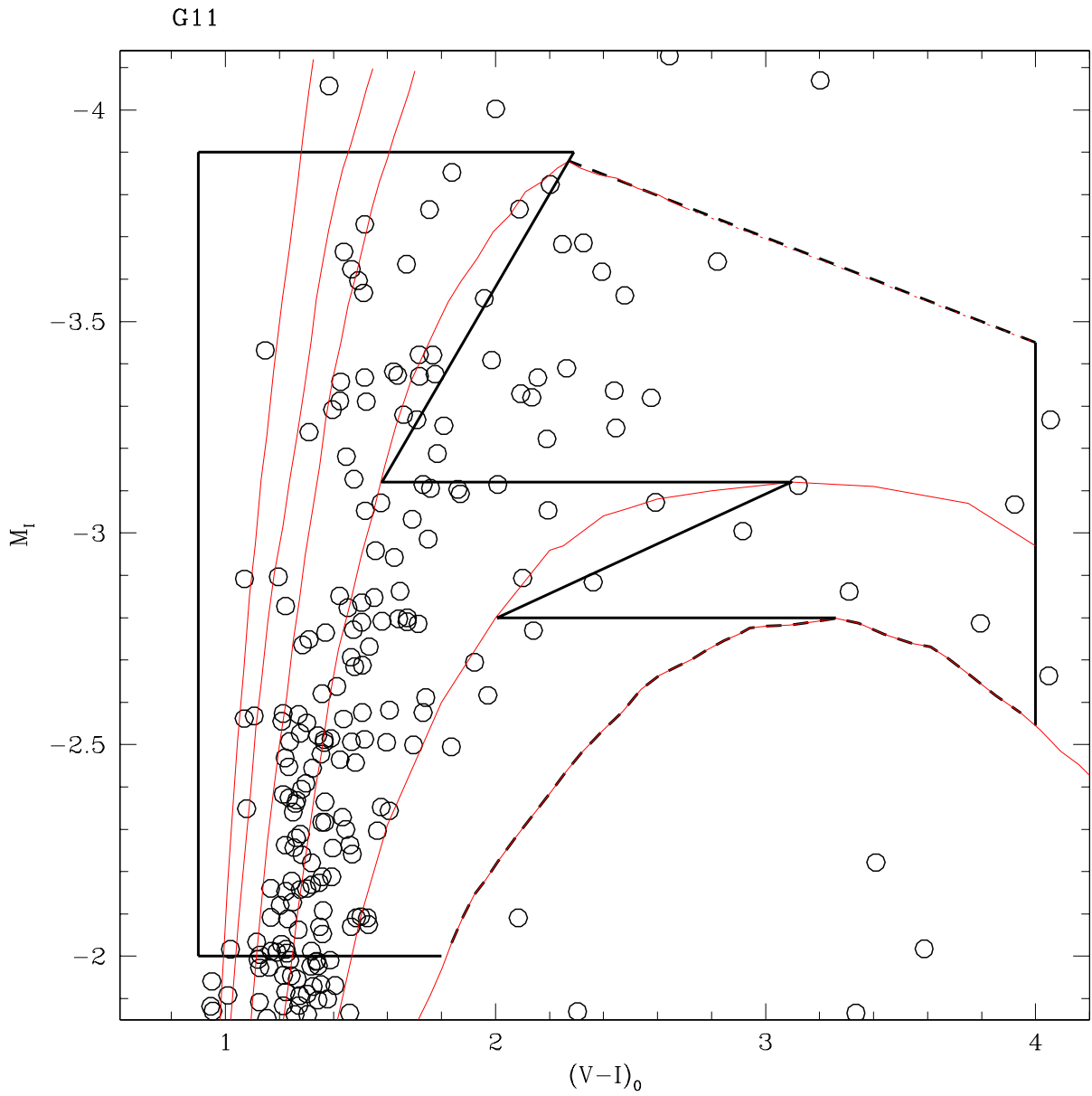}    
\caption{This plot illustrates how our interpolation scheme  
works.  To the left of jagged boundary, we interpolate in color, while  
to the right, metallicity is determined using an interpolation in magnitude.  
Stars falling below the curved locus of NGC 6528 are excluded from the sample,  
but represent only $\approx 1$\% of the total in any field. }   
\end{figure*}  
  
\clearpage   
  
\begin{figure*}  
\includegraphics[width=17cm]{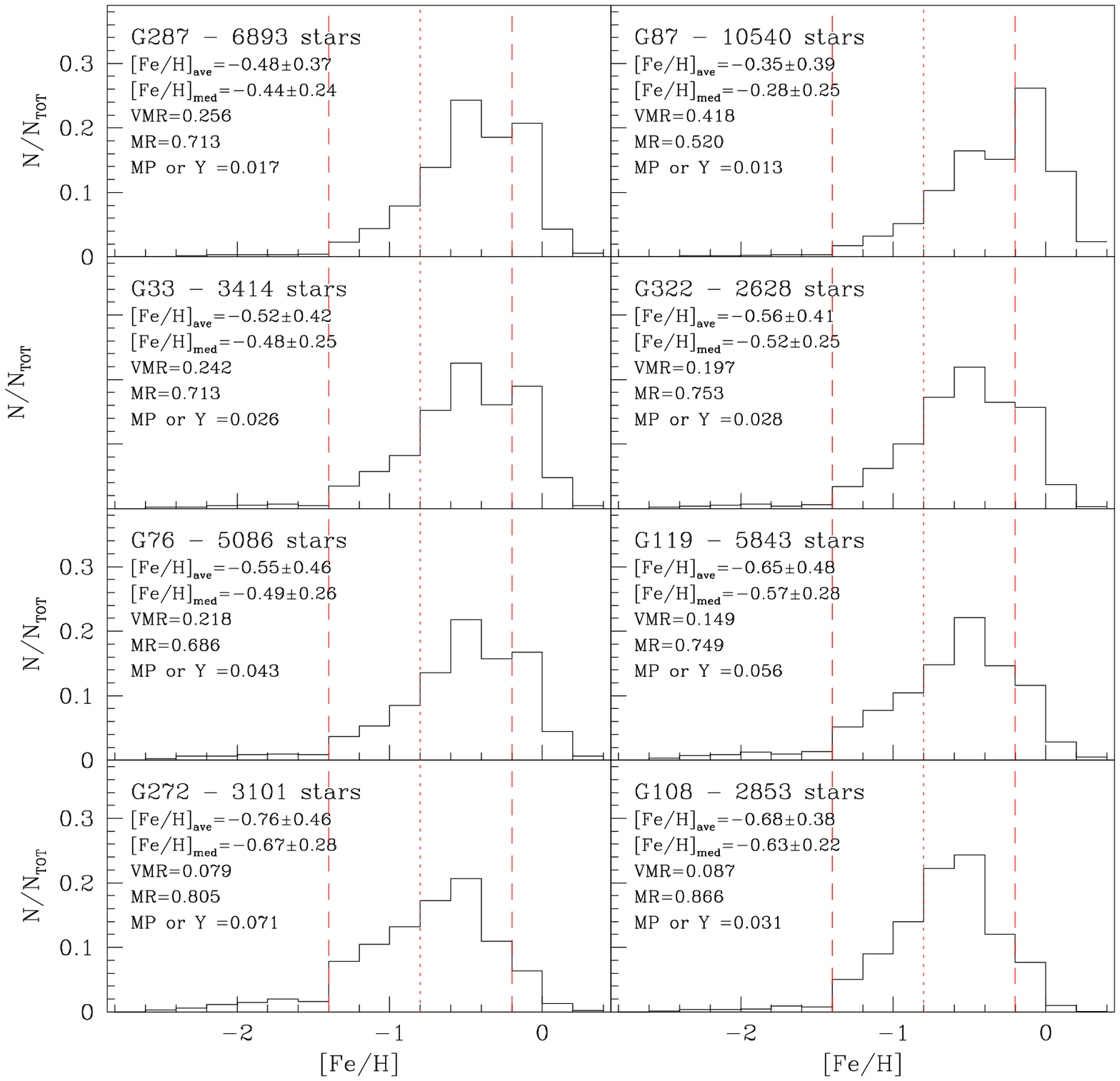}  
\caption{  
Histograms of the metallicity distributions, in the \citet{cg97} scale,  
for the fields: G287, G87, G33, G322, G76, G119, G272 and G108.  
In the upper left corner of each panel are reported: the name of the field, the  
number of stars used to derive the MD, the average metallicity ($[Fe/H]_{ave}$)  
together with the associated standard deviation, the median metallicity  
together with the associated semi-interquartile interval, and the fractions  
of stars with:   
$\rm [Fe/H]<-1.4$ [Metal-Poor or Young--MP or Y],  
$\rm -1.4<[Fe/H]<-0.2$ [Metal-Rich --MR],   
$\rm [Fe/H]>-0.2$ [Very Metal-Rich -- VMR], respectively.    
}   
\end{figure*}  
  
\clearpage  
  
\begin{figure*}  
\includegraphics[width=17cm]{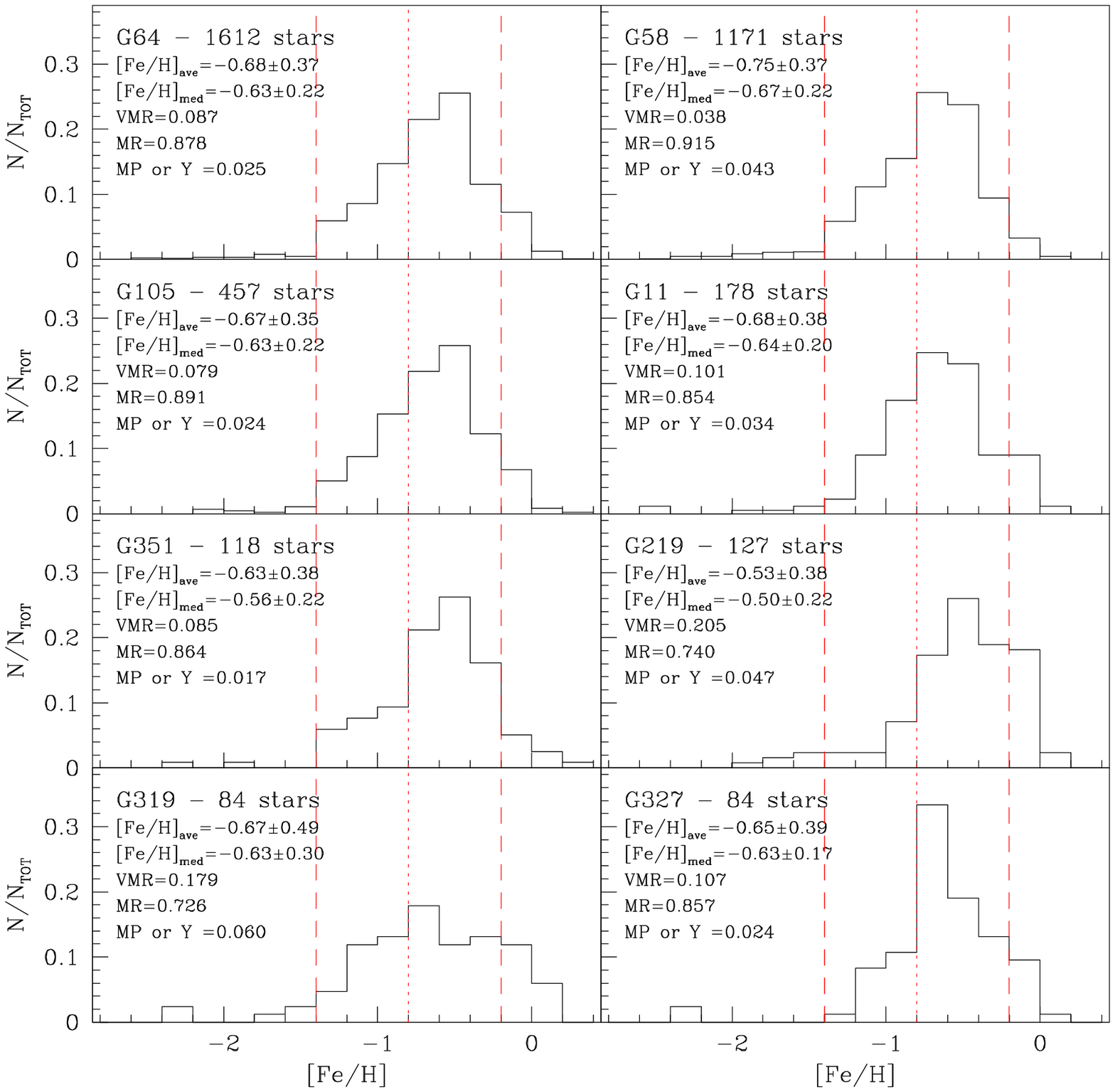}  
\caption{  
The same as Fig. 14 for the fields: G64, G58, G105,  
G11, G351, G219, G319 and G327.  
}   
\end{figure*}  
  
\clearpage  
  
\begin{figure*}  
\includegraphics[width=17cm]{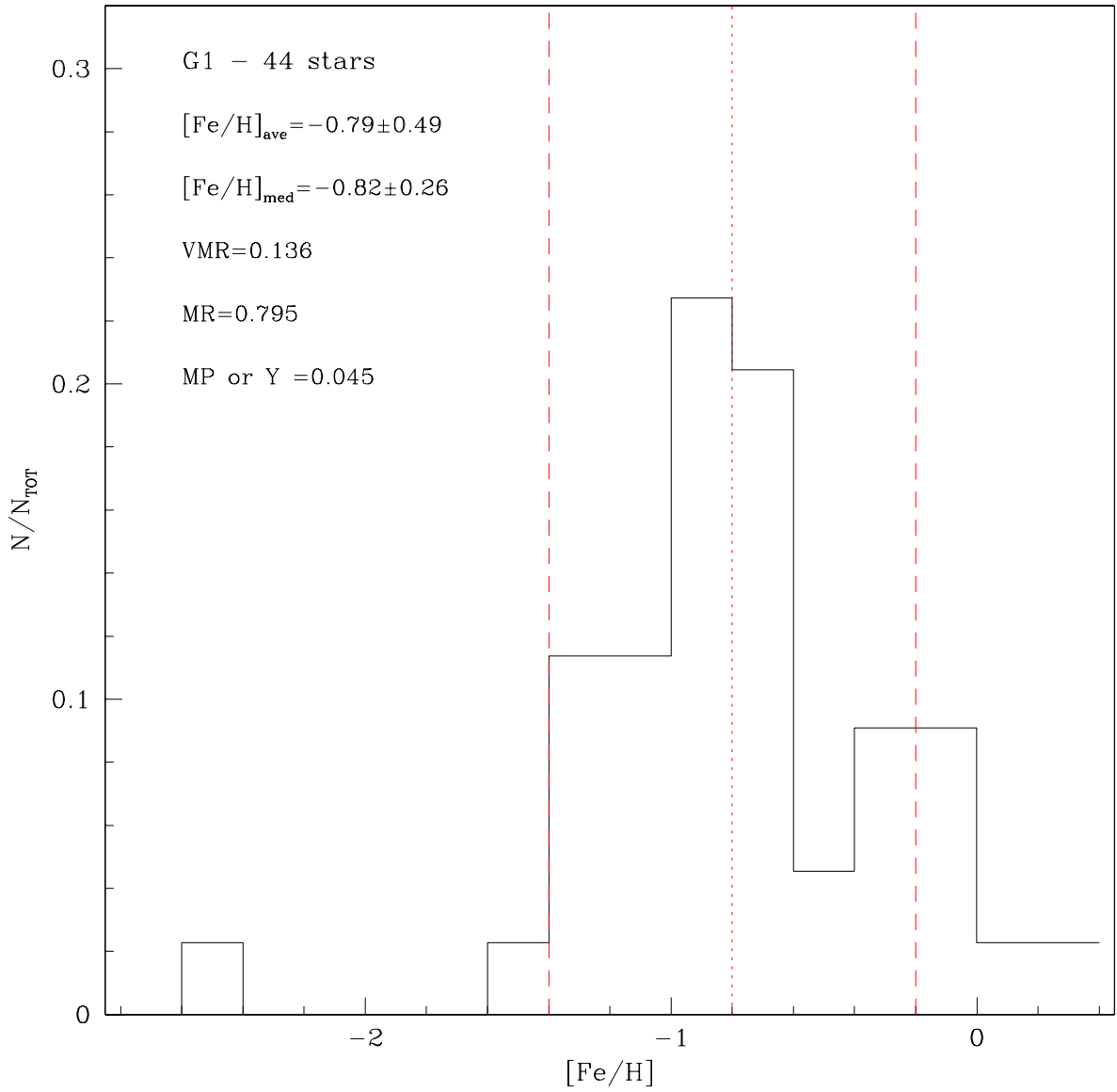}  
\caption{The metallicity distribution (same as Fig. 14, 15)  
for the field adjacent to the  
remote globular cluster G1, 35 kpc from the M31 nucleus.   
The frames from \citet{rich96} have been reduced in  
the same way as all other fields in our dataset. }   
\end{figure*}  
  
\clearpage  
  
\begin{figure*}  
\includegraphics[width=17cm]{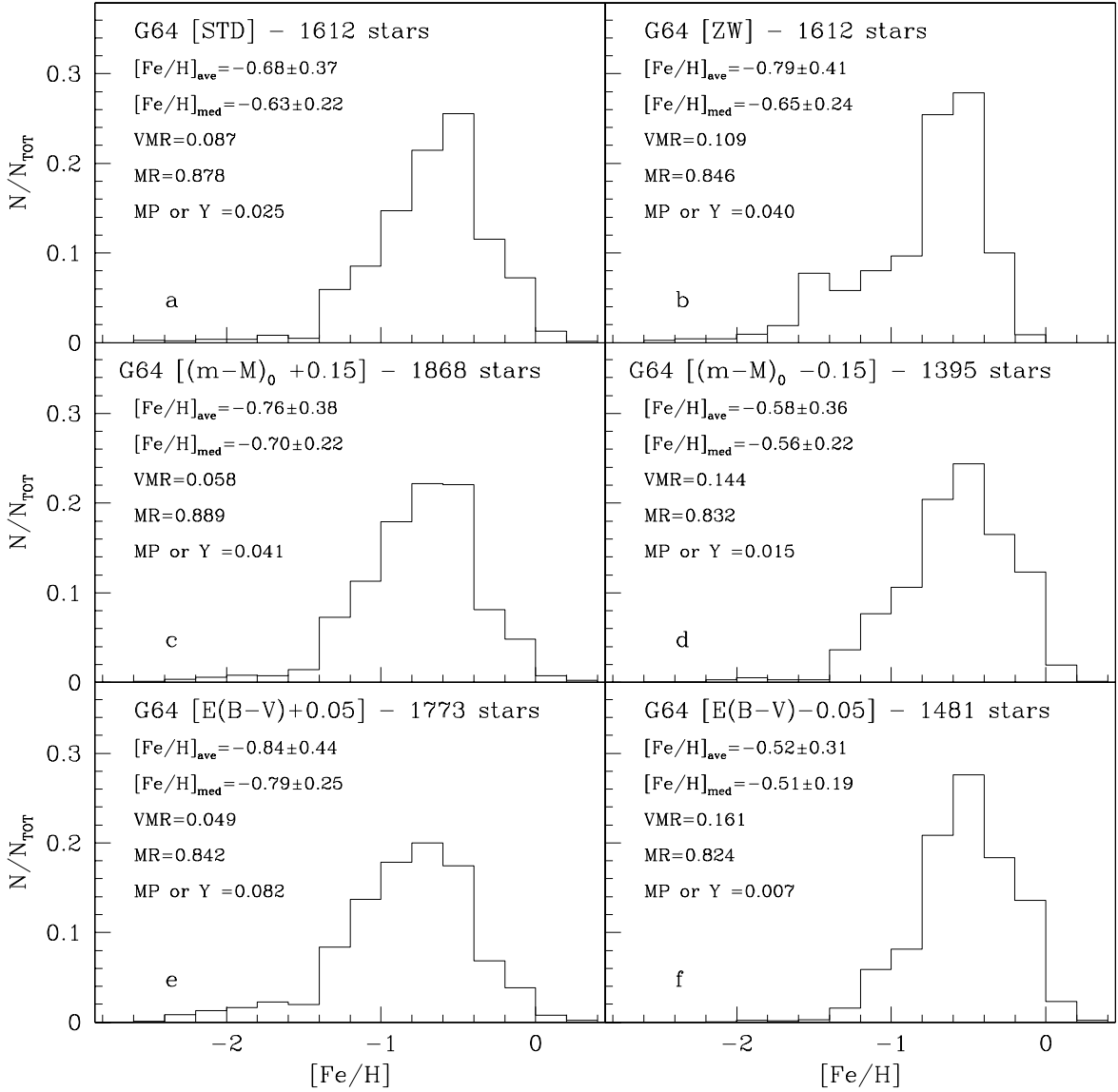}  
\caption{The metallicity distribution for the field G64   
shows the effect of choosing different parameters.  The detailed  
shape of the abundance distribution and the possible existence of  
a metal poor peak are both sensitive to these choices.  
Panel (a):  Using our adopted CG metallicity scale, distance modulus, and   
reddening. Panel (b): same as (a), ZW metallicity scale. Panel (c):   
$(m-M)_0$ increased by 0.15 mag, and (d): $(m-M)_0$ decreased   
by 0.15 mag. Panel (e): $E(B-V)$ increased by 0.05 mag, and (f): $E(B-V)$  
decreased by 0.05 mag.}   
\end{figure*}  
   
\clearpage  
  
\begin{figure*}  
\includegraphics[width=17cm]{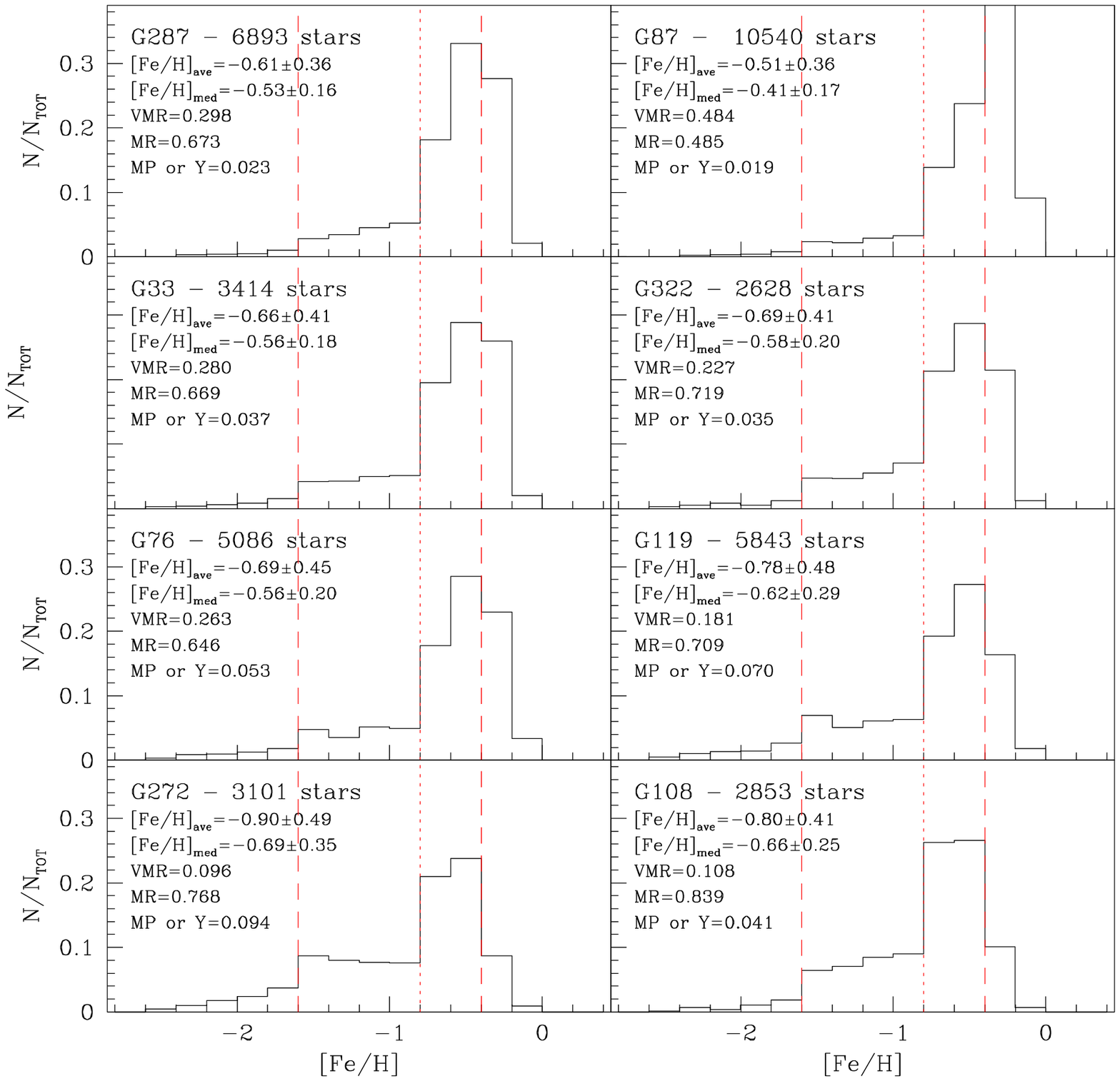}  
\caption{  
Histograms of the metallicity distributions, in the \citet{zw84} scale,  
for the fields: G287, G87, G33, G322, G76, G119, G272 and G108.  
In the upper left corner of each panel are reported: the name of the field, the  
number of stars used to derive the MD, the average metallicity ($[Fe/H]_{ave}$)  
together with the associated standard deviation, the median metallicity  
together with the associated semi-interquartile interval, and the fractions  
of stars with: $\rm [Fe/H]<-1.6$ [Metal-Poor or Young--MP or Y],  
$\rm -1.6<[Fe/H]<-0.4$ [Metal-Rich --MR], $\rm [Fe/H]>-0.4$ [Very   
Metal-Rich -- VMR], respectively.    
}    
\end{figure*}  
  
\clearpage   
  
\begin{figure*}  
\includegraphics[width=17cm]{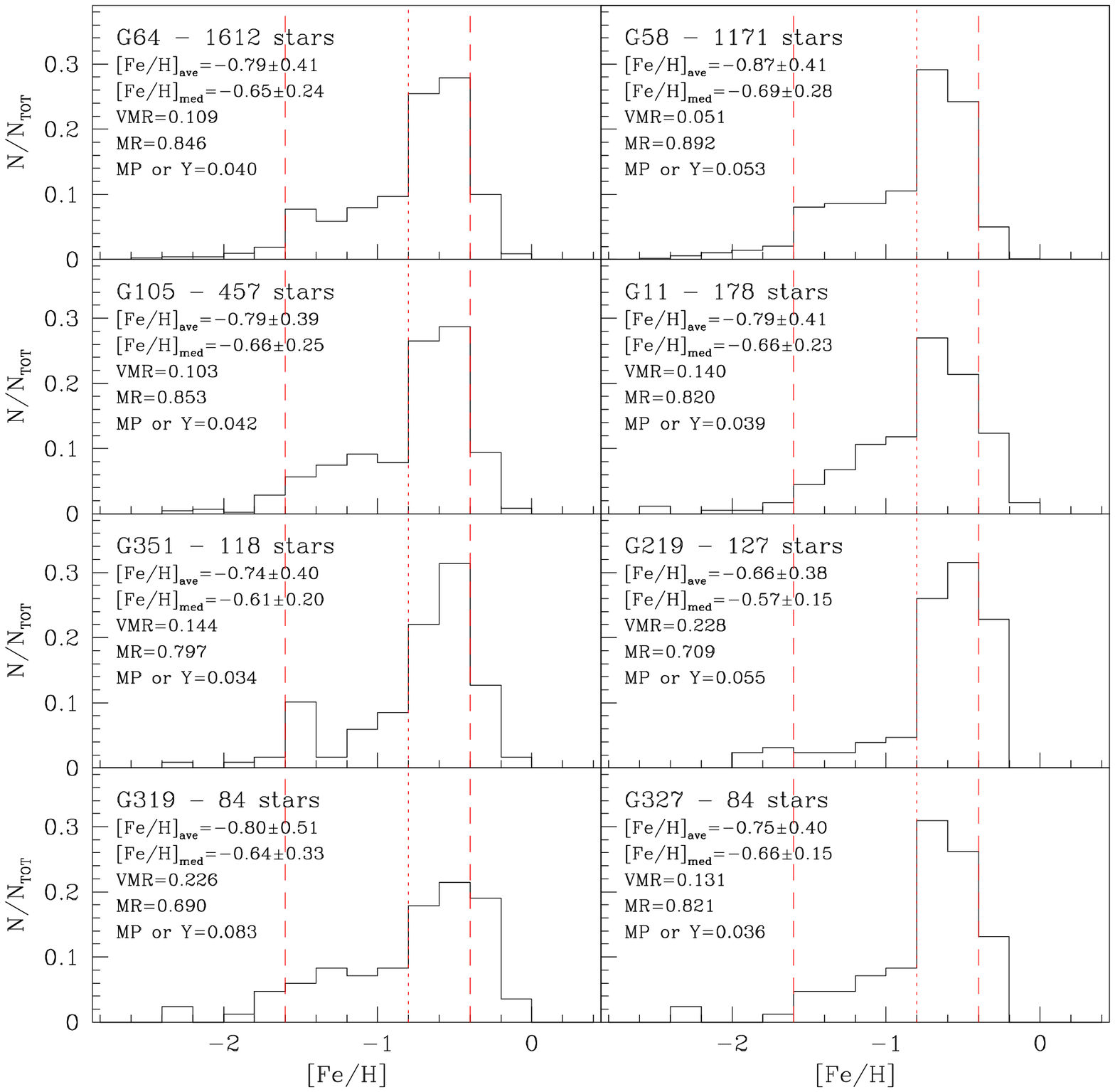}  
\caption{  
The same as Fig.  18 for the fields: G64, G58, G105,  
G11, G351, G219, G319 and G327.  
}    
\end{figure*}  
  
\clearpage   
  
\begin{figure*}  
\includegraphics[width=17cm]{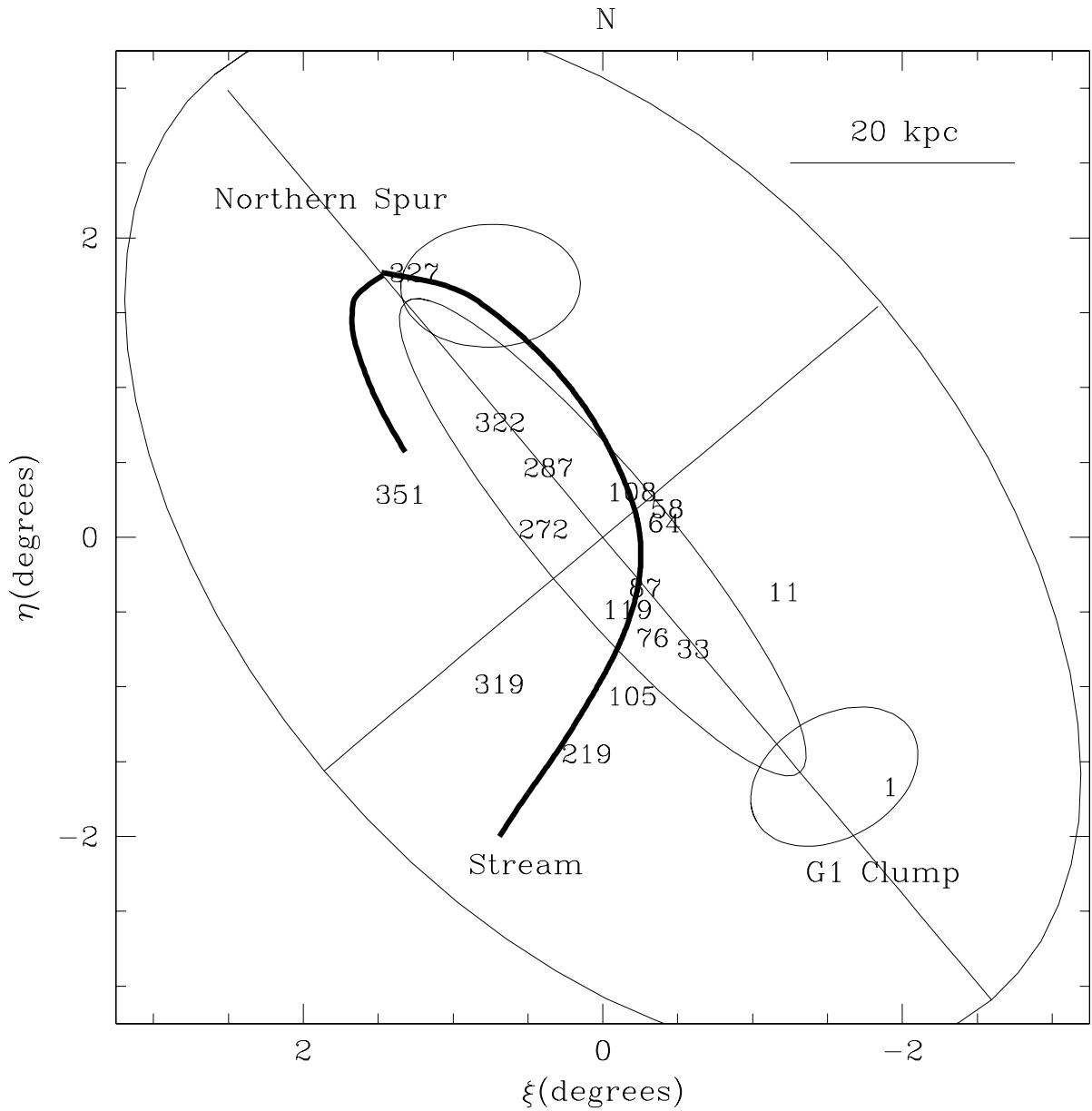}  
\caption{  
Map of the spatial and  chemical substructures   
as described by  \citet{fer02}  (their Fig. 7).  
The positions of our fields are plotted  
over the cartoon describing the location of the possible projected orbit  
of the "giant stellar stream", the "northern spur", and the  
"G1 clump".  
}    
\end{figure*}  
  
\clearpage   
  
\begin{figure*}  
\includegraphics[width=17cm]{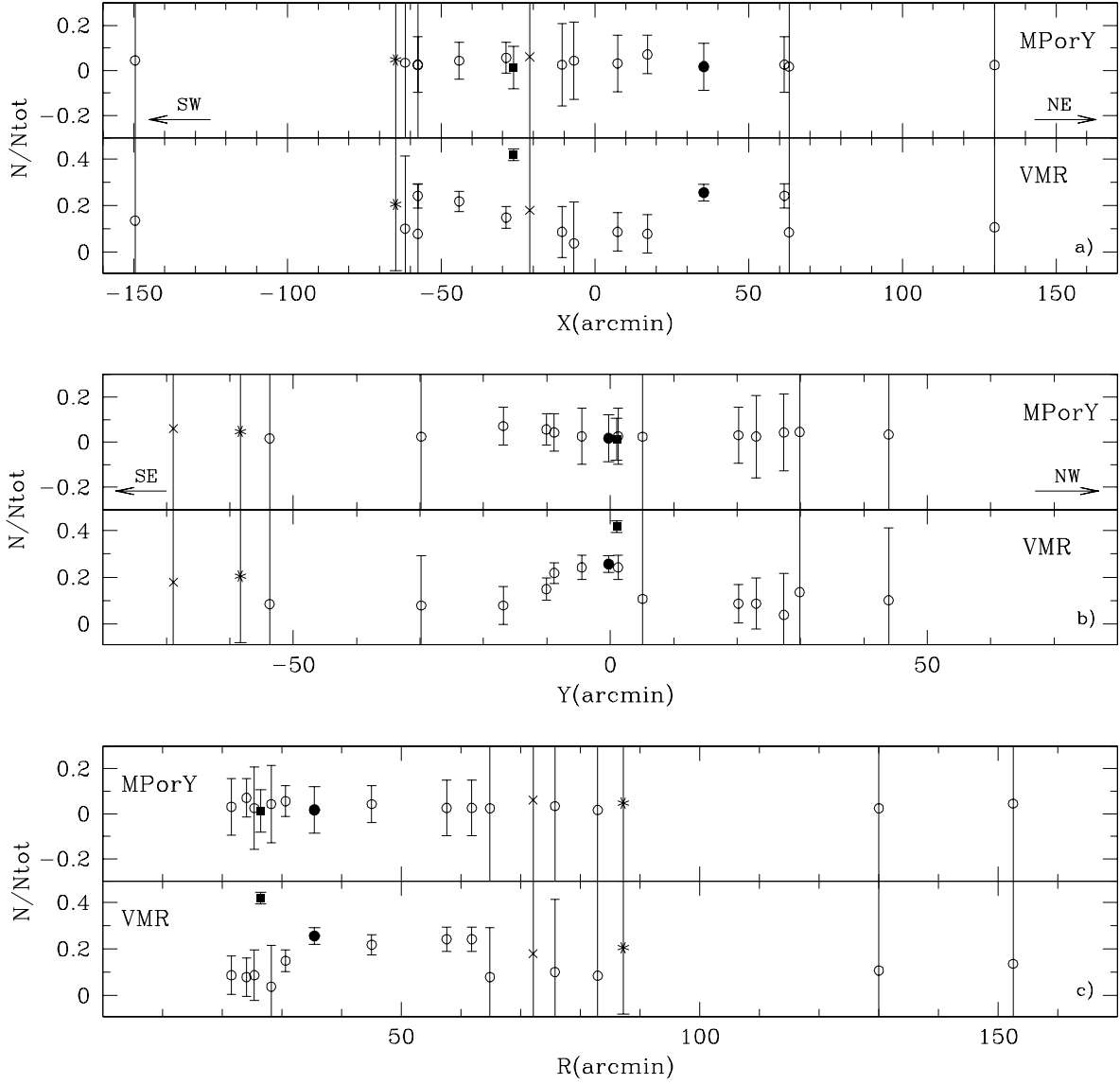}  
\caption{  
Metallicity parameters, derived from the MDs, are plotted  
versus the angular distance from the center of M31.  
$\rm [Fe/H]<-1.4$ [Metal-Poor or Young -- MP or Y], and  
$\rm [Fe/H]>-0.2$ [Very Metal-Rich -- VMR],     
versus (a) the X-coordinate, (b) Y, (c) R, the galactocentric  
distance, respectively. A few fields are evidenced by different 
symbols (see Sect. 9.2.1): G287 (solid circle); G87 (solid square);
G219 (star) and G319 ($\times$). }    
\end{figure*}

\clearpage   
  
\begin{figure*}  
\includegraphics[width=17cm]{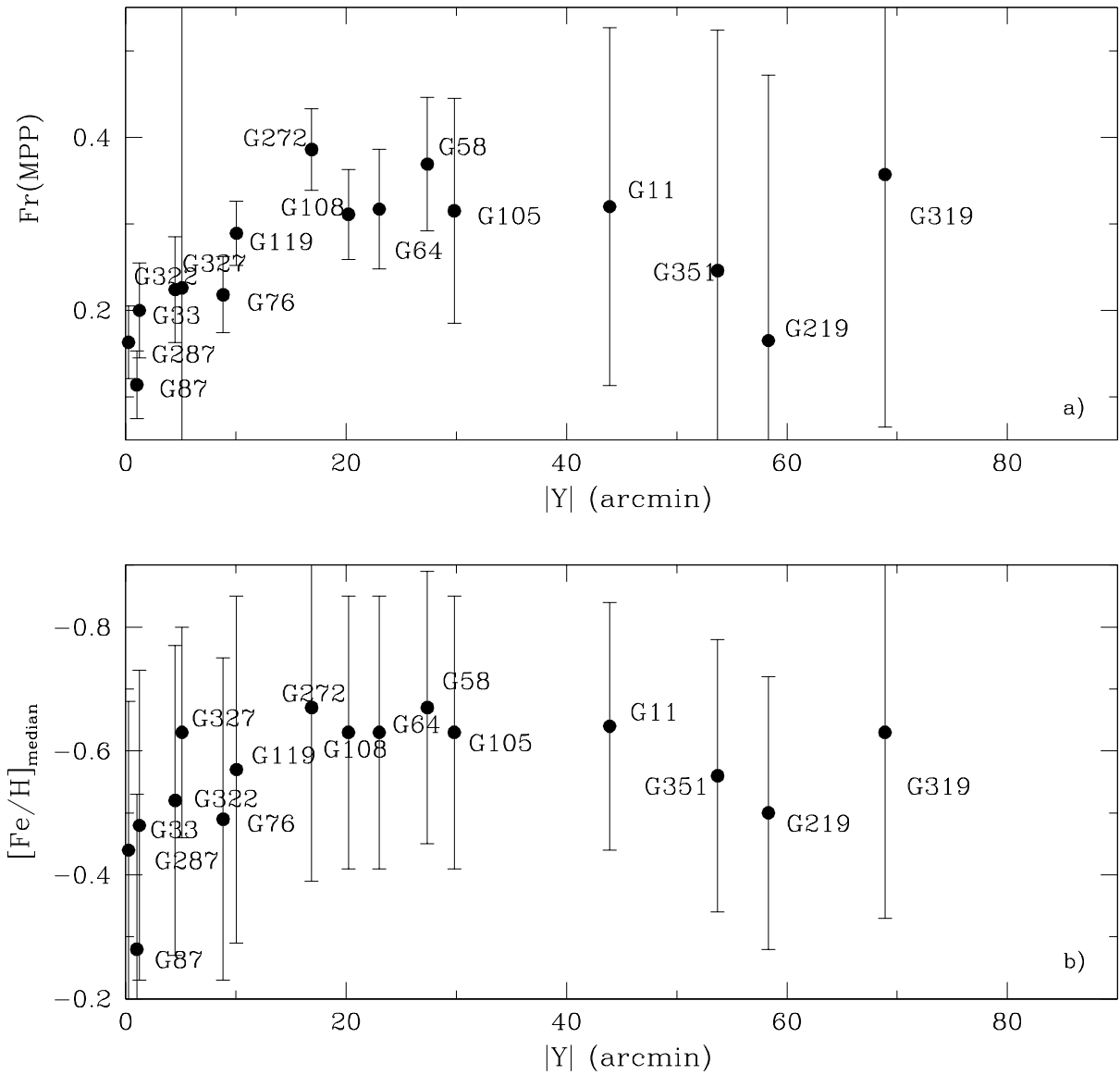}  
\caption{Panel (a): Fraction of the Metal-Poor Population (with   
$\rm [Fe/H]<-0.8$ [MPP] plotted versus the absolute distance  
from the major axis $ |Y|$. Panel (b): Median metallicity (GC97-scale)  
plotted vs. the absolute distance  from the major axis $ |Y|$.  
}   
\end{figure*}  

\end{document}